\documentclass{jfm}
\usepackage{color}

\usepackage{graphicx}
\usepackage{natbib}
\usepackage{psfrag}
\usepackage{amsmath}
\usepackage{mathrsfs}
\usepackage{subfigure}
\usepackage{amsfonts}
\usepackage[section]{placeins}
\usepackage{epstopdf}
\usepackage{float}
\usepackage{subfigure}
\usepackage{tabu}
\usepackage{booktabs} 

\usepackage{multirow}
\usepackage{amsmath}
\usepackage{amssymb}
\usepackage{stmaryrd}
\usepackage{color, psfrag,multicol}
\usepackage{upgreek}
\usepackage[percent]{overpic}

\newcommand*{\beq}{\begin {equation} }
\newcommand*{\eeq}{\end {equation} }
\newcommand*{\beqs}{\begin {equation*} }
\newcommand*{\eeqs}{\end {equation*} }

\title[A realizable data-driven approach to delay bypass transition with control theory]
{A realizable data-driven approach to delay bypass transition with control theory}

\author[P. Morra, K. Sasaki, A. Hanifi, A. V. G. Cavalieri, D. S. Henningson]%
{Pierluigi Morra$^{1}$,\ns
Kenzo Sasaki$^{2}$\break
Ardeshir Hanifi$^{1}$,\ns
Andr\'e V. G. Cavalieri$^2$,\ns
Dan S. Henningson$^1$}
\affiliation{$^1$KTH Royal Institute of Technology, Linn\'e FLOW Centre, SE-10044, Stockholm, Sweden\\[\affilskip]
$^2$Instituto Tecnol\'ogico de Aeron\'autica, S\~ao Jos\'e dos Campos, Brazil}

\volume{???}
\pagerange{???--???}
\date{?; revised ?; accepted ?. - To be entered by editorial office}
\begin{document}
\maketitle
\begin{abstract}
The current work presents a realizable method to control streaky disturbances in boundary layer flows and delay transition to turbulence by means of active flow control. Numerical simulations of the nonlinear transitional regime in a Blasius boundary layer are performed where streaks are excited in the boundary layer by means of a high level of free-stream turbulence. The occurring disturbances are measured by means of localized wall-shear-stress sensors and damped out using near-wall actuators, which resemble ring plasma actuators. Each actuator is powered by a time-varying signal whose amplitude is computed by processing signals from the sensors. The processed signal is the result of two control laws: the Linear Quadratic Gaussian regulator (LQG) and the Inverse Feed-Forward Control technique (IFFC). The use of the first control method, LQG, requires a state-space representation of the system dynamics, so the flow is described by means of a linear time-invariant operator that captures only the most relevant information of the dynamics and results in a reduced order model (ROM). The ROM is computed by means of the eigensystem realization algorithm (ERA), which is based on the impulse responses of the real system. Collecting such impulse responses may be unfeasible when considering free-stream turbulence because of the high dimensionality of the input forcing needed for a precise description of such a phenomenon. Here, a new method to identify the relevant system dynamics and generate the needed impulse responses is proposed, based on additional shear-stress measurements in an upstream location. Transfer functions between such measurements and other downstream sensors are obtained and allow the derivation of the ERA system, in a data-driven approach that would be realizable in experiments. Finally, the effectiveness of the technique in delaying bypass transition is shown. The work (i) presents a systematic and straightforward way to deal with high dimensional disturbances in order to build ROMs for a feasible control technique, and (ii) shows that even when considering practical constraints such as the type and size of actuators and sensors, it is possible to achieve at least as large delay of bypass transition as that obtained in more idealized cases found in literature.

\end{abstract}
\section{Introduction}
\label{sec:intro}
The laminar flow state is characterized by a lower friction drag than the turbulent one, which implies less energy consumption for many applications, such as transportation means like trains and aircrafts. Therefore, control of laminar-turbulent transition is of great interest in many technical areas.  The transition scenario depends on a number of parameters, and an overall picture of these different scenarios can be found in \cite{schmid2001a}. Transition to turbulence in boundary layer flows where free-stream turbulence has an intensity higher than $\approx 1\%$ occurs rapidly and bypasses the classical scenario triggered by Tollmien-Schlichting (TS) waves, as showed by \cite{arnal1978a}. When free-stream turbulence is present, a set of low-frequency vortices (\cite{hultgren1981a}; \cite{hunt1999a}; \cite{zaki2009a}) enters the boundary layer and causes the appearance of elongated streaky structures of alternating high and low streamwise velocity. This was firstly observed in the experimental studies of \cite{klebanoff1971a}. The amplitude of such velocity fluctuations grows linearly along the streamwise direction \citep{andersson1999a,luchini2000a} and is accompanied by growing secondary fluctuations of the streaky structures on the planes perpendicular to the streamwise direction. When the amplitude of such secondary cross-flow fluctuations is sufficiently high turbulent spots appear \citep{brandt2002a}, which grow and merge further downstream and ultimately lead to a fully turbulent flow. This process was observed both in experiments \citep{matsubara2001a} and simulations \citep{brandt2004a}. Thus, the boundary layer can be divided into three zones: (i) an upstream zone where there is high level of receptivity and free-stream turbulence triggers disturbances in the boundary layer, (ii) a middle zone where streaky disturbances grow due to the linear lift-up mechanism, and (iii) a downstream zone where the flow nucleates turbulent spots which grow and merge as they propagate downstream until the boundary layer becomes fully turbulent.\\ \indent
The boundary layer flow in the middle zone can often be described with sufficient accuracy by the linearized Navier-Stokes equations. The possibility to work with a linear system greatly facilitates the application of flow control techniques. The a priori knowledge of the linear behavior of the TS-waves  was exploited in the experiments of \cite{thomas1983a} and in the simulations of \cite{laurien1989a} to counteract TS-waves and delay transition. Similarly, for bypass transition, \cite{jacobson1998a} exploited the linearity of the dynamical system to show the possibility to damp streaky structures. In those works the a priori knowledge of the system dynamics was used to create ad hoc counter disturbances. Such ad hoc practice lacks in generality and may require tedious testing, therefore it is appealing to apply the optimal control theory to flow control problems. The control-theory community has produced many reliable and elegant techniques to tackle linear systems. Among the first successful applications of the optimal control theory in fluid mechanics are the works of \cite{joshi1997a}, \cite{bewley1998a}, \cite{hogberg2000a} and \cite{hogberg2003a}, where optimal control methods are applied to linearized systems and used in fully non-linear channel flows. More recently, \cite{monokrousos2008a} showed the successful application of the Linear Quadratic Gaussian regulator (LQG) for control of streaks triggered by the free-stream turbulence.\\ \indent
In optimal control techniques the final goal is finding the function that takes measurements as input and gives actuation signals as output while minimizing an objective function. Particularly, in classical optimal control methods the optimal solution for linear time-invariant systems is given by solving an algebraic Riccati equation, which consists of a matrix equation whose dimensions are roughly as those of the original linear system to control. If the original linear system has large dimensions, as is the case in fluid mechanics, the solution of the algebraic Riccati equation may be extremely computationally demanding. A possible solution is reducing the order of  the optimal control problem by keeping only the information useful for the control. This is the idea behind reduced order models (ROMs). In fact, measurements usually contain only a portion of the total information present in the system and actuators can usually excite only certain structures. In control theory, such limitations posed by sensors and actuators define two properties of the system: its observability and its controllability, respectively. The control problem alone needs only the portion of the system that is observable and controllable. The practice of model reduction in flow control was treated in \cite{bagheri2009c}, \cite{semeraro2011a}, \cite{poussot2015a} and \cite{yao2017a}. The approach was shown to be successful in the sense that the solution to the control problem was nearly unaffected by the use of a ROM. A classic technique for achieving a ROM is the Eigensystem Realization Algorithm (ERA) \citep{juang1985a,semeraro2013a}. ERA is based on a set of impulse responses from each input (actuators and disturbances) to each output (measurements).\\ \indent
\cite{ma2011a} showed that the ROM achieved by the ERA is equivalent to that achieved by balanced truncation. This means that the ROM resulting from the ERA is a projection of the original system onto the set of modes given by the intersection of the set of the most observable flow structures and the set of the most controllable flow structures. Qualitatively, an \textit{observable} structure is one that generates non-zero outputs whereas a \textit{controllable} structure is one that can be excited by the inputs. The term ``most" is obviously case dependent. For controllable structures it represents the number of flow structures used to recreate with an acceptable small error the flow field obtained by an impulse response. The same reasoning holds for the observable flow structures but with respect to the adjoint system (see \cite{bagheri2009c} for a more detailed discussion of controllability and observability in fluid mechanics systems). Examples of ERA applications in fluid mechanics are found in  \cite{semeraro2013a}, for control of a three-dimensional non-linear TS-wave packet, and \cite{sasaki2018a}, for control of three-dimensional TS-waves arising from stochastic disturbances.\\ \indent
A consistent modeling of the inputs implies correctly modeling the space spanned by the disturbances, which may require the use of a basis with as many degrees of freedom as the dimensions of the desired space. Thus, in case the space spanned by a disturbance or an actuator has large dimensions it may become unfeasible to collect all the impulse responses. A similar issue may also happen when the number of outputs is very high. Another possibility to avoid demanding computations for solving the control problem is dropping the use of model-based methods as discussed by \cite{fabbiane2014a}, who made use of a learning algorithm that needs only minimal modeling.\\ \indent
The present work addresses the delay of bypass transition in a physically realizable framework. We use a finite number of localized near-wall actuators that resemble ring plasma actuators \citep{kim2016a,kim2016b,shahriari2018a} and localized wall-shear-stress sensors. We also assess behavior of two model-based optimal control methods: the LQG regulator and the Inversion Feed Forward Control (IFFC) \citep{sasaki2018a}.  Moreover, we make use of the ERA to generate the ROM, and present a technique that allows to account for the free-stream turbulence without resorting to the same high dimensional basis used for the description of the disturbance in the fully non-linear simulations. This permits us to collect a smaller set of impulse responses, which makes the system associated with the control problem much smaller and less computationally demanding. This technique is based on measurements only, which makes it feasible in experiments as well.\\ \indent
The paper is structured as follows. In \S~\ref{sec:goveqs} the equations used to describe the full or reduced system dynamics are introduced; in \S~\ref{sec:description_control} the control techniques of interest are briefly described; in \S~\ref{sec:plant} the details of the framework for the non-linear simulations are outlined; in \S~\ref{sec:modelling_techniques} the identification techniques and the used identified models are presented; in \S~\ref{sec:transdelay} the behavior of the designed controller in the non-linear Navier-Stokes is assessed. A summary of  the main conclusions is given in \S~\ref{sec:concl}.\par
%
%
\section{Governing equations}\label{sec:goveqs}
\subsection{Dynamical system}
The Navier-Stokes equations can be expressed in terms of the perturbation quantities as
\begin{subequations}\label{eq:NSsys00}
\begin{align}
&\frac{\partial \mathbf{q'}}{\partial t} = -(\mathbf{q'} \cdot \nabla) \mathbf{q}_B-(\mathbf{q}_B \cdot \nabla) \mathbf{q'}-(\mathbf{q'} \cdot \nabla) \mathbf{q'} -\nabla p' + Re^{-1} \nabla^2\mathbf{q'} + \mathbf{f},\label{eq:NSsys00a}\\
&\nabla \cdot \mathbf{q'}=0,\\
&\mathbf{q'} = \mathbf{q'}_0 \quad \text{at} \quad t=0,
\end{align}
\end{subequations}
where $\mathbf{q'}= \mathbf{q'}(\mathbf{x},t)$ is the perturbation velocity vector, $\mathbf{q}_B=\mathbf{q}_B(\mathbf{x})$ the unperturbed velocity vector, $p'=p'(\mathbf{x},t)$ the perturbation pressure, $\mathbf{f}=\mathbf{f}(\mathbf{x},t)$ a body force vector, $Re$ the Reynods number, $t$ the time, $\mathbf{x}=(x_1,x_2,x_3)^T$ the space variable, and $\nabla = (\partial_{x_1},\partial_{x_2},\partial_{x_3})$ the gradient operator. \\ \indent
Here, the unperturbed velocity vector $\mathbf{q}_B$ is the solution of an evolving zero-pressure-gradient boundary layer.
The velocity perturbation $\mathbf{q'}$ satisfies no-slip conditions at the wall $x_2 = 0$ and Neumann conditions at free-stream $x_2 = L_{x_2}$. Periodicity is assumed along the spanwise direction $x_3$ and enforced along the streamwise direction $x_1$ by means of an artificial forcing $\mathbf{f}_{BC} = \lambda(x_1) \mathbf{q}'$, which is placed in a fringe region at the outlet. $\lambda(\mathbf{x})$ is a non-negative function which is non-zero only within the fringe region. $\mathbf{f}_{BC}$ forces all perturbations to zero and modifies $\mathbf{q}_B$ to be periodic (see \cite{nordstrom1999a} for details about $\mathbf{f}_{BC}$).\par
The flow control problem consists in finding the correct external action that modifies the fluid dynamics to achieve a specific goal. In our case, such external action can take the form of a boundary condition or a body force and can be expressed as a function of time and space. It follows that the problem can be split into finding the correct spatial distribution of such an action and its time modulation. In the present work it is assumed that the spatial distribution and the time modulation of the external action are decoupled. The spatial distribution is prescribed, so the flow control problem reduces to the computation of its time-varying amplitude. From now on this time-varying scalar is referred to as \textit{input}. Using a finite number of actuators $N_u$, the external action used for control reads
\begin{equation}\label{eq:extact}
\mathbf{f}_{u} =  \sum_{k=1}^{N_u} u(t)_k\mathbf{b}(\mathbf{x})_k,
\end{equation}
where $\mathbf{b}(\mathbf{x})_k$ is the spatial shape of the $k$-th body force, and $u(t)_k$ the corresponding time variation. The latter represents the control input.\par
Free-stream turbulence is modeled as a forcing in the fringe region. $\mathbf{f}_{BC}$ is modified to force $\mathbf{q}'$ to be equal to a prescribed perturbation that mimics the presence of free-stream turbulence. The prescribed perturbation is of the form
\begin{equation}
\mathbf{q}'_{FST} = \sum_{\alpha} \sum_{\beta} \sum_{\omega} \Phi(\alpha,\beta,\omega) \hat{\mathbf{q}}'(x_2,\alpha,\beta,\omega) e^{i(\alpha x_1 + \beta x_3 - \omega t)},
\end{equation}
with $\hat{\mathbf{q}}'$ an eigensolution to the Orr-Sommerfeld Squire eigenvalue problem for a parallel flow in a semi-bounded domain, $\alpha$ the streamwise wavenumber, $\beta$ the spanwise wavenumber, and $\omega$ the angular frequency. Free-stream disturbances are thus expanded as a sum of eigenfunctions of the linearized parallel-flow problem (see \cite{brandt2004a} for more details).\par
A linearized version of the Navier-Stokes equations about $\mathbf{q}_B$ can be obtained by dropping the non-linear term $(\mathbf{q'} \cdot \nabla) \mathbf{q'}$ from equation \eqref{eq:NSsys00a}. \par
%
\subsection{Reduced-order dynamical system}
\begin{figure}%
\begin{center}
{\includegraphics[width=0.97\textwidth,trim={120 300 220 110},clip]{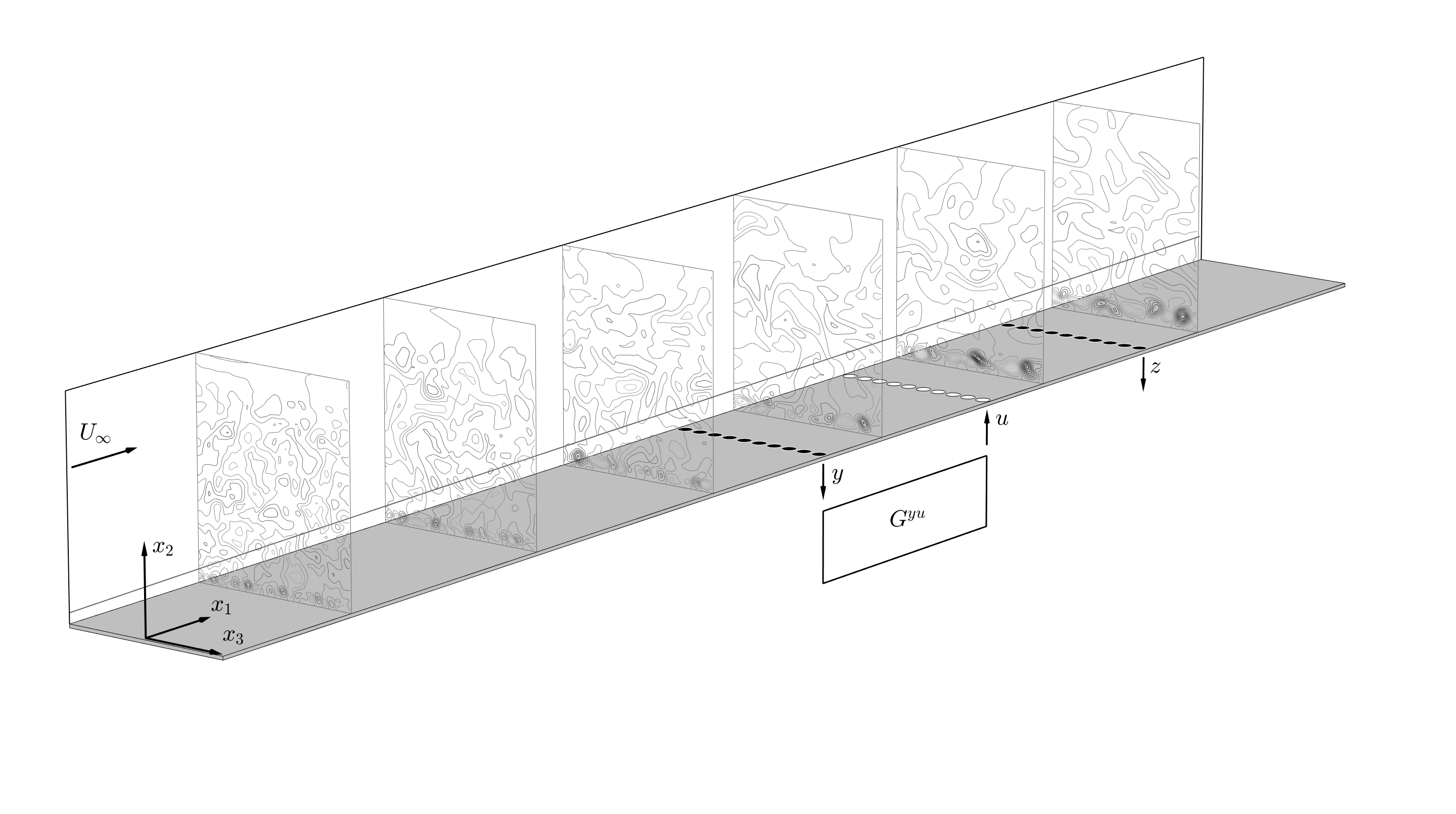}}\\
\end{center}
 \caption{Plant. Computational box (cut along $x_1$): frame of reference $x_1x_2x_3$, sensors $y$ and $z$ (black circles), actuators $u$ (white circles), and controller $G^{yu}$. Contour plots: perturbation part of the streamwise velocity, $q_1'$, snapshot of an uncontrolled case at time $t=t^*$; the isolines of the contour plots are all for the same set of values. Boundary layer, $\delta_{99}$, shown on the left wall of the box.}
\label{fig:plant}
\end{figure}

The linear dynamical system used for the application of control theory techniques is a ROM and reads
\begin{equation}\label{eq:NSsys01}
\mathbf{\dot{q}}(t) = \mathbf{A} \mathbf{q}(t) + \mathbf{B} \mathbf{u}(t) + \mathbf{M}_d \mathbf{d}(t),
\end{equation}
where $\mathbf{q}= \mathbf{q}(t)$ is the $N\times1$ state vector (which generally is not exactly the same quantity represented by $\mathbf{q'}$), $\mathbf{\dot{q}}$ is its time derivative, $\mathbf{A}$ the $N\times N$ matrix that defines the system dynamics, $\mathbf{B}$ the $N\times N_u$ matrix that characterizes the control inputs, $\mathbf{u}=\mathbf{u}(t)$ a $N_u \times 1$ column vector containing all the input amplitudes $u(t)_k$, $\mathbf{M}_d$ the $N\times N_d$ matrix that characterizes the disturbance inputs, and $\mathbf{d}=\mathbf{d}(t)$ a $N_d \times 1$ column vector containing all the input amplitudes $d(t)_k$. $N$ is the degree of freedom of the ROM, $N_u$ the number of control inputs, and $N_d$ the number of disturbance inputs.\\ \indent
We also assume to have access to two finite sets of measurements: $\mathbf{y}(t)$, $N_y \times 1$, and $\mathbf{z}(t)$, $N_z \times 1$, where $N_y$ and $N_z$ represent the respective number of measurements available. It holds
\begin{equation}
\mathbf{y}(t) = \mathbf{C}_y \mathbf{q}(t), \quad \mathbf{z}(t) = \mathbf{C}_z \mathbf{q}(t),
\end{equation}
where the $N_y \times N$ matrix $\mathbf{C}_y$ and the $N_z \times N$ matrix $\mathbf{C}_z$ characterize the measurements in the ROM. From now on $\mathbf{y}(t)$ and $\mathbf{z}(t)$ are referred to as \textit{outputs}. Equations (2.4) and (2.5) form a ROM state-space representation of the system.\\ \indent
A different description of the system can be given by means of \textit{transfer functions} (TF). TFs are built by performing the Laplace transform on the state-space representation and in general describe the system as function of the angular frequency $\omega$ only. Here, sensors and actuators are placed on straight lines along the spanwise direction (figure~\ref{fig:plant}), with $N_u = N_y = N_z$. Since the flow is periodic in the spanwsie direction, TFs, inputs and outputs can be expressed as functions of the spanwise wavenumber $\beta$ as well. The description by means of TFs reads
\begin{equation}\label{eq:TFsys03}
\begin{aligned}
\hat{y}(\omega,\beta_k) &= \hat{G}^{uy}(\omega,\beta_k) \hat{u}(\omega,\beta_k) + \hat{G}^{dy}(\omega,\beta_k) \hat{d}(\omega,\beta_k),\\
\hat{z}(\omega,\beta_k) &= \hat{G}^{uz}(\omega,\beta_k) \hat{u}(\omega,\beta_k) + \hat{G}^{dz}(\omega,\beta_k) \hat{d}(\omega,\beta_k),\\
\end{aligned}
\end{equation}
where $\hat{G} = \hat{G}(\omega,\beta_k)$ are the TFs, $\hat{y} = \hat{y}(\omega,\beta_k)$ and $\hat{z}=\hat{z}(\omega,\beta_k)$ the outputs in the frequency domain, $\hat{u} = \hat{u}(\omega,\beta_k)$ and $\hat{d}=\hat{d}(\omega,\beta_k)$ the inputs in the frequency domain, and $k$ is used to stress the fact that the number of outputs is finite, so there is a finite amount of available wavenumbers. From now on all the variables denoted by a hat symbol are function of $(\omega,\beta_k)$, and the explicit writing $(\omega,\beta_k)$ is dropped.\\ \indent
The description of the system by means of TFs can be translated in the physical domain, leading to
\begin{equation}\label{eq:FIRsys00}
\begin{aligned}
y(t)_k = \int_0^t \sum_{m=1}^{N_u} G^{uy}_{km}(t-\tau) u(\tau)_m \ \mathrm{d}\tau + \int_0^t \sum_{m=1}^{N_d} G^{dy}_{km}(t-\tau) d(\tau)_m \ \mathrm{d}\tau,\\
z(t)_k = \int_0^t \sum_{m=1}^{N_u} G^{uz}_{km}(t-\tau) u(\tau)_m \ \mathrm{d}\tau + \int_0^t \sum_{m=1}^{N_d} G^{dz}_{km}(t-\tau) d(\tau)_m \ \mathrm{d}\tau,\\
\end{aligned}
\end{equation}
with $k$ the output index, and $m$ the input index.\par
%
%
\section{Control techniques}\label{sec:description_control}
The present configuration of outputs and inputs together with the convective nature of the flow make all the control techniques described in this section be in a feed-forward configuration \citep{belson2013a}.\par
The control techniques used in the present work are all based on the assumption that the input $u(t)_k$ is a function of the upstream outputs $y(t)_m$,
\begin{equation}\label{eq:y2u}
u(t)_k = \int_0^t \sum_{m=1}^{N_y} G^{yu}_{km}(t-\tau) y(\tau)_m \ \mathrm{d}\tau,
\end{equation}
with $ k = 1,2,\dots,N_u$. Since $\mathbf{q}_B$ is independent of the spanwise direction $x_3$, the instantaneous linearized system dynamics is homogeneous along $x_3$. The latter and the fact that the outputs are all given by the same type of sensor allows to drop the usage of the index $k$ in \eqref{eq:y2u} to have $G^{yu}_{m}$. Then, \eqref{eq:y2u} can be rewritten as
\begin{equation}\label{eq:y2u00}
u(t)_k = \int_0^t \sum_{m=1}^{N_y} G^{yu}_{m}(t-\tau) y(\tau)_{m+k-1} \ \mathrm{d}\tau,
\end{equation}
where for $m+k-1 > N_y$ spanwise periodicity implies the use of $m+k-1 - N_y$.\par
%
\subsection{Linear Quadratic Gaussian regulator}
The technique is based on a linear model, aims at minimizing a quadratic cost function, and assumes the presence of Gaussian white noise disturbances.\\ \indent
The addition of Gaussian white noise on the output $\mathbf{y}(t)$ in the state-space representation reads
\begin{equation}\label{eq:LQGsys00}
\begin{aligned}
\mathbf{\dot{q}} (t)&= \mathbf{A} \mathbf{q}(t) + \mathbf{B} \mathbf{u}(t) + \mathbf{M}_d \mathbf{d}(t),\\
\mathbf{y}(t) &= \mathbf{C}_y \mathbf{q}(t) + \mathbf{n}(t) ,\\
\mathbf{z}(t)&= \mathbf{C}_z \mathbf{q}(t),\\
\end{aligned}
\end{equation}
where $\mathbf{n}(t)$, $N_y \times 1$, is the time modulation of the noise. $\mathbf{d}(t)$ is also treated as white noise. There is no addition of noise on the output $\mathbf{z}(t)$ because it represents a reference output to minimize, whose measurement is not available in reality. The noise on the output $\mathbf{y}(t)$ corresponds to noise in a real available measurement.\\ \indent
The covariance matrices associated with $\mathbf{d}(t)$ and $\mathbf{n}(t)$ are $\mathbf{V}_d$, $N_d \times N_d$, and $\mathbf{V}_n$, $N_y \times N_y$, respectively, and are both diagonal and constant because of the assumption that $\mathbf{d}(t)$ and $\mathbf{n}(t)$ are white noise disturbances; in particular, it can be written
\begin{equation}\label{eq:covmat}
\mathbf{V}_d = v_d \mathbf{I},\quad \quad \mathbf{V}_n = v_n \mathbf{I},
\end{equation}
with $v_d > 0$ and $v_n > 0$ real scalars and $\mathbf{I}$ the identity matrix.\\ \indent
The technique consists in finding $G^{yu}_m$ by minimizing a prescribed $\mathcal{H}_2$-norm of interest. The disturbances $\mathbf{d}(t)$ and $\mathbf{n}(t)$ are treated as random variables, so the objective function of interest is defined as the expected value of an $\mathcal{H}_2$-norm. Here, the objective function contains both the reference output $\mathbf{z}(t)$ and the input for the control $\mathbf{u}(t)$, which is added to avoid an infinite amplitude of the input signal, penalizing excessive control action. The objective function reads
\begin{equation}
\label{eq:functionalforlqg}
J = \mathbb{E} \left[ \lim_{T\to\infty} \frac{1}{T}\int_0^T \mathbf{z}(t)^T \mathbf{Q} \mathbf{z}(t) + \mathbf{u}(t)^T \mathbf{Ru}(t) \ \mathrm{d}t \right],
\end{equation}
where the $N \times N$ matrices $\mathbf{Q}$ and $\mathbf{R}$ are design weights. The operator $\mathbb{E}[\bullet]$ represents the expected value. From now on the matrices $\mathbf{V}_d$, $\mathbf{V}_n$, $\mathbf{Q}$ and $\mathbf{R}$ are referred to as weight matrices or design weights.\\ \indent
In the LQG it is assumed that $\mathbf{u}(t)$ is a linear function of the states, but it is also assumed that not all the states are known at each time instant, so a second system for state estimation is introduced. The estimation system makes use of the known outputs to reconstruct the states at each instant of time, and is designed to minimize the estimation error. Thus, in addition to the minimization of the objective function \eqref{eq:functionalforlqg} to compute the input that controls the system, the estimation introduces a second minimization problem. Generally these two minimization problems are coupled, but in the LQG they are independent and solved separately. They consists in the Linear quadratic regulator, which solves the control problem by assuming full-state information, and the Kalman filter, which solves the estimation problem by assuming stochastic disturbances on the outputs. The solution of the LQG is the combination of the two independent solutions. More details about the LQG are given in Appendix~\ref{app:LQG}, whereas a thorough description can be found in \cite{lewis1995a}.\\ \indent
%
\subsection{Inversion Feed Forward Control}
Inversion feed forward control (IFFC) is a technique developed in the frequency domain, and is based on a system described by TFs \eqref{eq:TFsys03}. The technique is exactly the same one used in \cite{sasaki2018a}, but in that work it is referred to as Wave Cancellation. The authors decided to change the nomenclature to adopt the name used in the control community.\\ \indent
The contribution of the disturbance $\hat{d}$ in the second equation of \eqref{eq:TFsys03} may also be expressed as
\begin{equation}\label{eq:TFobs00}
\hat{G}^{dz}\hat{d} = \hat{G}^{yz} \hat{y} + \hat{p},
\end{equation}
where $\hat{G}^{yz}$ is a TF to design in order to maximize the extraction of information from $\hat{y}$, while $\hat{p}$ is the residual part of the information in $\hat{z}$ which is not retrieved by $\hat{G}^{yz}\hat{y}$. The loss of information, i.e. $\hat{p} \neq 0$, may be unavoidable and can be seen by resorting to the state-space representation. The matrix $\mathbf{C}_y$ that characterizes the output $\mathbf{y}(t)$ does not necessarily span the same space spanned by the matrix that describes the system dynamics $\mathbf{A}$, so the outputs $\mathbf{y}(t)$, being in a subspace, cannot reconstruct the whole state space. It follows that the only portion of the signal $\hat{z}(\omega,\beta_k)$ which can be obtained from the outputs $\mathbf{y}(t)$ is
\begin{equation}\label{eq:TFobs01}
\tilde{\hat{z}} = \hat{G}^{uz} \hat{u} + \hat{G}^{yz} \hat{y},
\end{equation}
where $\tilde{\hat{z}}$ is an estimate of $\hat{z}$.\\ \indent
The objective of the control problem is the annihilation of the output $\tilde{\hat{z}}$. Then, a straightforward strategy to solve the problem is imposing $\tilde{\hat{z}} = 0$, which is the basic idea behind IFFC. Assuming $\hat{u} = \hat{K}\hat{y}$ in \eqref{eq:TFobs01} gives
\begin{equation}\label{eq:IFFCsol00}
\hat{K} = (\hat{G}^{uz})^{-1}\hat{G}^{yz},
\end{equation}
where $\hat{K}$ solves the control problem in the frequency-wavenumber domain. The result in \eqref{eq:IFFCsol00} is ill-conditioned in the zeros of $\hat{G}^{uz}$, which may lead to spurious high amplitudes of the input. Moreover, model uncertainties are not considered. Such limitations are addressed in \cite{devasia2002a}, where the TFs, the inputs and the outputs are functions of the angular frequency $\omega$ only. Here, the same approach is used with some modification to account for a system description as function of $\omega$ and $\beta_k$, following \cite{sasaki2018a}. The technique makes use of two weights, $\hat{R}$ and $\hat{Q}$, which here are taken as constant, and solves the control problem by minimizing the following prescribed objective function
\begin{equation}
\label{eq:WCobj00}
J =\int_{-\infty}^{\infty}\sum_k \left(\hat{u}^{H}\hat{R}\hat{u}+\hat{z}^{H}\hat{Q}\hat{z}\right) \Delta \beta_k \ \mathrm{d}\omega,
\end{equation}
where the superscript $H$ indicates the complex conjugate transpose. The presence of the objective function turns the nature of the problem into an $\mathcal{H}_2$ optimal control problem, whose solution is given by
\begin{equation}
\label{optimizedkernel}
\hat{K}=\frac{(\hat{G}^{uz})^H\hat{Q}\hat{G}^{yz}}{\hat{R}+(\hat{G}^{uz})^H\hat{Q}\hat{G}^{uz}}.
\end{equation}
The inverse Fourier transform of $\hat{K}$ gives $G^{yu}_m$ as in equation \eqref{eq:y2u00}; only the causal part of $G^{yu}_m$ is used for control, since actuation must be decided based solely on present or past information from the sensors..\\ \indent
This technique was already applied by \cite{sasaki2018a,sasaki2018b} and showed to be successful in the control of Kelvin-Helmholtz and Tollmien-Schlichting waves, where the equivalence between LQG and IFFC for the damping of TS waves was also shown.\par
%
%
\section{Plant}\label{sec:plant}
The domain of interest is a box as shown in figure~\ref{fig:plant}, where the white symbols represent the outputs and the black symbols the inputs. For flow simulations the pseudo-spectral code SIMSON \citep{chevalier2007a} is used. Here, the reference length is taken to be the displacement thickness of the boundary layer at the inlet $\delta_0^*$ and the reference velocity is the free-stream velocity $U_{\infty}$. In all of the present simulations the Reynolds number is $Re = U_{\infty} \delta_0^*/\nu = 300$. All the results that involve transition to turbulence are performed by means of LES simulations on a box of dimensions $(L_{x_1},L_{x_2},L_{x_3}) = (4000,60,50)$ with $(N_{x_1},N_{x_2},N_{x_3}) = (1024,121,108)$ points for the discretization. The effect of the large-eddy simulations (LES) filter (see \cite{schlatter2004a}, \cite{schlatter2006a}, \cite{schlatter2006b} for details) in the area where the flow dynamics is linear is negligible \citep{monokrousos2008a}. All the results that do not need to include the fully turbulent regime are performed by direct-numerical simulations (DNS) on a box of dimensions $(L_{x_1},L_{x_2},L_{x_3}) = (1000,60,50)$ with $(N_{x_1},N_{x_2},N_{x_3}) = (1152,121,108)$ points for the discretization. The points along the wall-parallel directions are equi-spaced, whereas along the wall-normal there are Gauss-Lobatto points.\\ \indent
The free-stream turbulence is modeled by superposition of 200 random modes from the continuous Orr-Sommerfeld-Squire spectrum. The integral length scale and the turbulent intensity used for all presented results are respectively $L = 7.5 \delta_0^*$ and $Tu = 3.0\%$, considering the free-stream turbulence spectrum in \cite{brandt2004a}.\\ \indent
As shown in figure~\ref{fig:plant}, the input and output devices are placed along straight lines. The first set of outputs is placed at $x_{1,y} = 250$, which is downstream of the zone with high receptivity. The second set of outputs is placed at $x_{1,z} = 400$, since after that position the non-linearities start to be non-negligible. Input signals, corresponding to actuators, are generated at $x_{1,u} = 325$ to have the same $\Delta x_1$ between input and outputs, such that the traveling time of a disturbance from the first set of outputs to the inputs is roughly the same to the traveling time from the inputs to the second set of outputs. The chosen location for the devices is also optimal in terms of identification accuracy for control design, as shown in Appendix \ref{app:pos}. The number of devices along the spanwise direction is the same for each set and it is equal to $N_u = N_y = N_z = 36$. Such a choice is motivated by analyzing the wavenumber spectrum of the average disturbance energy. According to Shannon information theorem the sampling wavenumber needs to be at least twice the wavenumber of interest. In this case measuring the highest non-negligible spanwise fluctuation would require at least 18 devices. In order to have a better measurement of the spanwise fluctuations 36 devices are used. The devices are equi-spaced along the spanwise direction. The shape of the input actuator is given by
\begin{equation}
\mathbf{b}(\mathbf{x}) = \{0, b_2(\mathbf{x}), 0\}^T,
\end{equation}
with
\begin{equation}
b_2(\mathbf{x}) = \exp\left[-\frac{(x_1-x_{1,0})^2}{\sigma_{x_1}^2}-\frac{(x_2)^2}{\sigma_{x_2}^2}-\frac{(x_3-x_{3,0})^2}{\sigma_{x_3}^2}\right],
\end{equation}
where $\sigma_{x_1} = 3$,  $\sigma_{x_2} = 5$, and $\sigma_{x_3} = 1.5$. The actuator shape resembles that of ring plasma actuators (see \cite{shahriari2018a,kim2016a,kim2016b}), generating a body force in the wall-normal direction. This is efficient to excite or cancel streaks due to the lift-up effect. A detailed analysis on the effect of the actuator shape is presented in \cite{sasaki2019a}, which is the parallel work to the present one.\\ \indent
The outputs are computed as
\begin{equation}
\frac{1}{S} \int_S \frac{\partial q'_1}{\partial x_2} \bigg\rvert_{x_2 = 0} \ \mathrm{d}S,
\end{equation}
with $S$ the area on the wall where the measure is taken. This is an averaged measure of the shear stress associated to the perturbation part of the streamwise velocity component on the wall.\par
%
%
\section{Reduced-order modelling and control design}
\label{sec:modelling_techniques}
Both control methods introduced in \S~\ref{sec:description_control} are model-based. The IFFC technique requires the knowledge of two TFs, $\hat{G}^{zu}$ and $\hat{G}^{yz}$. $\hat{G}^{zu}$ is by definition the Fourier transform of the output signal resulting from an impulse-response simulation of the linearized Navier-Stokes, whereas $\hat{G}^{yz}$ needs to be modeled. The LQG technique, instead, requires the knowledge of the matrices $\mathbf{A},\ \mathbf{B},\ \mathbf{C}_{y},\ \mathbf{C}_{z}$ and $\mathbf{M}_d$, which characterize the ROM and need to be modeled.\\ \indent
The techniques used for this modeling are introduced in the remainder of this section, and are all based on input-output data, which is usually available in experiments. In input-output data part of the information about the system dynamics is lost. However, its usage is a reasonable design choice, since the control techniques work only with the observable and controllable structures, whose time evolution is described by input-output signals. In fact, by definition, the information lost in input-output data is the one associated to the unobservable and uncontrollable structures. \\ \indent
The present configuration of outputs and inputs together with the convective nature of the flow allow to estimate the downstream outputs $\mathbf{z}(t)$ from the upstream outputs $\mathbf{y}(t)$. This fact is exploited in the following part of this section.\par
%
\subsection{Empirical TFs}\label{subsec:empiricaltf}
The estimation of downstream outputs $\hat{z}$ by means of upstream outputs $\hat{y}$ can be performed by designing a TF $\hat{G}^{yz}$. Here, $\hat{G}^{yz}$ is computed by means of an identification technique using the information extracted from the output data. The TF obtained in this way is referred to as \textit{empirical} TF. This method was introduced in \cite{sasaki2018c} for the estimation of a turbulent jet and applied in \cite{sasaki2018b} for the closed-loop control of a two-dimensional shear layer. Here, the approach is extended to a flow with spanwise periodicity, i.e. outputs are function of $\beta_k$ as well. It was shown in \cite{bendat2011a} that the optimal frequency response, in the least square sense, is defined from the auto- and cross-spectra of the input and output signals
\begin{equation}
\label{eq:empiricaltf}
\hat{G}_{yz}=\frac{\hat{S}_{yz}}{\hat{S}_{yy}},
\end{equation}
where $\hat{S}_{yy}$ and $\hat{S}_{yz}$ are respectively the auto- and cross-spectra of the input and output signals. Both $\hat{S}_{yy}$ and $\hat{S}_{yz}$ are computed as the expected values of $\hat{y}^H\hat{y}$ and $\hat{y}^H\hat{z}$, which are obtained via the process of ensemble averaging \citep{bendat2011a}. Equation \eqref{eq:empiricaltf}, sometimes referred to as an $\hat{H}_1$ estimator \citep{rocklin1985a}, minimizes the error due to noise in the output.\\ \indent
One desirable property of an $H_1$ estimator is that the prediction error is linearly uncorrelated to the available output signal \citep{rocklin1985a,bendat2011a}. Any remaining errors correlated to the available signal are either due to the presence of noise in the measurements or to spectral leakage, which is unavoidable because the signal is not exactly periodic in time. Spectral leakage can be minimized by using long time series, by windowing the signal for the ensemble averaging, or via the calculation of an improved frequency response, as outlined in the following section.\par
%
\subsection{Improved frequency response}\label{subsec:improvedtf}
The method considered here is referred to as \textit{improved} frequency-response and consists of improving the accuracy of a TF by means of an iterative algorithm. This allows to obtain a more accurate linear approximation of a system and is particularly interesting when the impulse responses of the disturbances are not available or unfeasible to collect, as is the case for the free-stream turbulence or experimental implementations. The method is designed to minimize noise, spectral leakage and capture some nonlinearity \citep{schoukens1997a}.\\ \indent
The algorithm is initialized with a first-guess TF $\hat{G}^{yz}_{0}$, which may be, for instance, the result obtained from equation \eqref{eq:empiricaltf}. Then, the estimation error, which is the difference between the signal obtained by using $\hat{G}^{yz}_{0}$ and the available output $\hat{y}$,  is computed. The error reads
\begin{equation}\label{eq:impTFerr00}
e(t)_k = z(t)_k - \int_0^t \sum_m G^{yz}_{0,m}(t-\tau) y(t)_m \mathrm{d} \tau.
\end{equation}
\noindent with $G^{yz}_{0,m}(t)$ the inverse Fourier transform of $\hat{G}^{yz}_{0}$. Then, the TF between the error and the available output $\hat{y}$ is computed as $\hat{G}^{yz}_{e}=\hat{S}_{ye}/\hat{S}_{yy}$, which is used to update the initial TF as $\hat{G}^{yz}_{1}=\hat{G}^{yz}_{0}+\hat{G}^{yz}_{e}$. Iterations are performed until the error TF is minimized.\par
%
\subsection{Eigensystem realization algorithm using TFs}\label{subsec:era}
In \S~\ref{sec:description_control} it was shown that in order to design the LQG regulator it is necessary to solve two algebraic Riccati equations, and the computational power required for their solution grows quickly with the dimensions of the matrix $\mathbf{A}$, which is the linear time-invariant operator used to describe the linearized system dynamics. Clearly, for fluid mechanics systems, which in general present numerous degrees of freedom, the usage of a ROM is preferable \citep{kim2007a}.\\ \indent
For the realization of the ROM we use the ERA \citep{juang1985a}. The ERA is based on output signals resulting from impulse responses. It is necessary to have access to an amount of impulse responses equal to the number of total inputs of the systems. The signals are written in a Hankel matrix whose dimensions depend on the total number of inputs and outputs and on the length of the saved time series, i.e. $N_t(N_y+N_z) \times N_t(N_u+N_d)$, $N_t$ being the number of time samples needed to have a good representation of the impulse response. In case the disturbance is free-stream turbulence the number of degrees of freedom used for the implementation of $\mathbf{d}(t)$ in the fully non-linear Navier-Stokes solver is of the order of hundreds. Then, since the Hankel matrix is decomposed by the Singular Value Decomposition, it is clear that collecting such a high number of impulse responses results in a heavy computational problem. Besides, in a practical application it is not possible to collect the impulse responses from the free-stream turbulence disturbance.\\ \indent
Therefore, in order to reduce the computational power required and to have a method that can be applied in experiments as well, a different approach is proposed. A new set $N_y \times 1$ of outputs $\mathbf{y}_{d}(t)$, which measure the same quantity as $\mathbf{y}(t)$ and $\mathbf{z}(t)$, is introduced upstream of $\mathbf{y}(t)$, and the outputs generated by a non-linear Navier-Stokes simulation with free-stream turbulence and without control action are stored. An impulse response coincides with a TF by definition, so the following TFs can be computed as in equation \eqref{eq:empiricaltf},
\begin{equation}\label{eq:d2yz}
\hat{G}^{y_{d}y}=\frac{\hat{S}_{y_{d}y}}{\hat{S}_{y_{d}y_{d}}},\quad \quad
\hat{G}^{y_{d}z}=\frac{\hat{S}_{y_{d}z}}{\hat{S}_{y_{d}y_{d}}},
\end{equation}
and their inverse Fourier transforms can be used as a set of impulse responses to mimic the presence of free-stream turbulence upstream every control device. These estimated impulse responses are used in the ERA to model the impulse responses coming from $\mathbf{M}_d\mathbf{d}(t)$ in equation \eqref{eq:LQGsys00}, which represents the disturbance in the system. The number of impulse responses from the actuators, which are characterized by $\mathbf{Bu}(t)$, is $N_u$. Nevertheless, only one of these impulse responses is collected because the other ones can be computed by exploiting the homogeneity of $\mathbf{q}_B$ and the periodicity of the flow along the spanwise direction.\\ \indent
Once the whole set of impulse responses is available it is possible to build the mentioned Hankel matrix, apply the ERA, and retrieve the ROM needed for the design of the LQG. To the best of the authors' knowledge, this is the first time this approach is used in fluid mechanics.\par
%
\subsection{Identified reduced order model}
\begin{figure}
\begin{center}
{\includegraphics[width=0.49\textwidth,trim={29 8 30 10},clip]{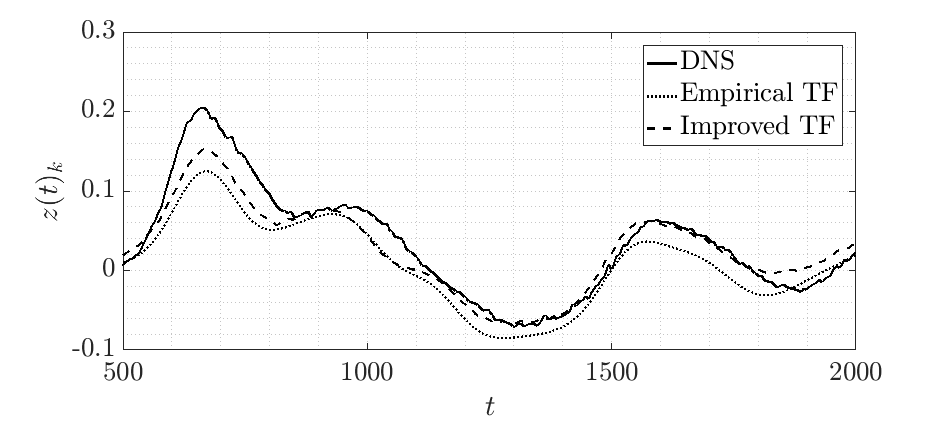}}
{\includegraphics[width=0.49\textwidth,trim={29 8 30 10},clip]{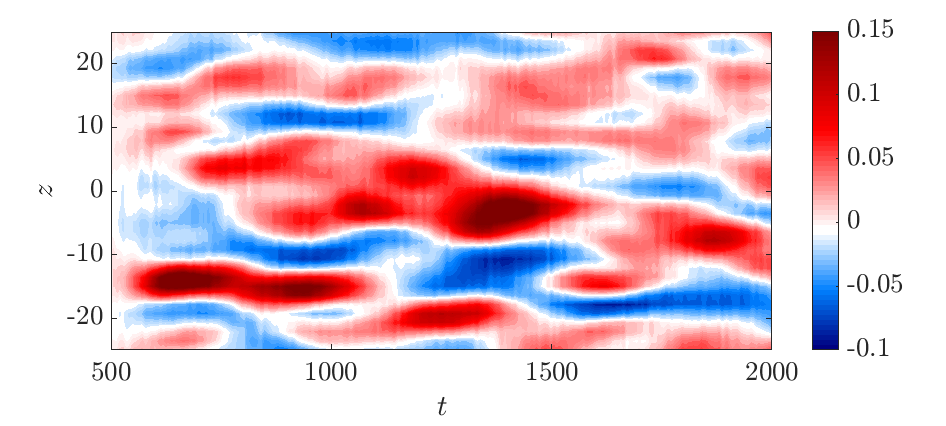}}
{\includegraphics[width=0.49\textwidth,trim={29 8 30 10},clip]{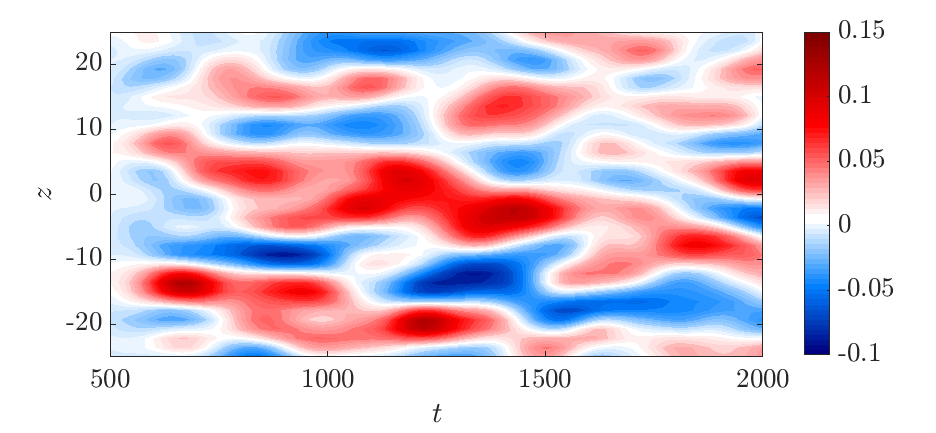}}
{\includegraphics[width=0.49\textwidth,trim={29 8 30 10},clip]{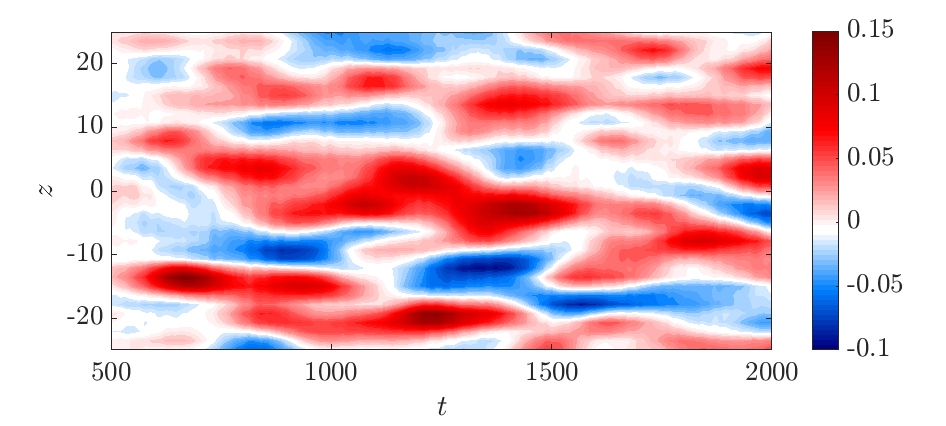}}
\end{center}
\caption{Comparison between the true $z(t)_k$ output from DNS data at $x_1=400$ and the estimated output $\tilde{z}(t)_k$. The available output $y_{avail}(t)_k$ is at $x_1=250$. Top-Left: single output $k=9$. Top-Right: DNS data $z(t)_k$. Bottom-Left: Estimated data $\tilde{z}(t)_k$, empirical TF. Bottom-Right: Estimated data $\tilde{z}(t)_k$, improved TF.}
  \label{fig:imprtf}
\end{figure}

\begin{figure}  
\begin{center}
{\includegraphics[width=0.48\textwidth,trim={110 50 160 20},clip]{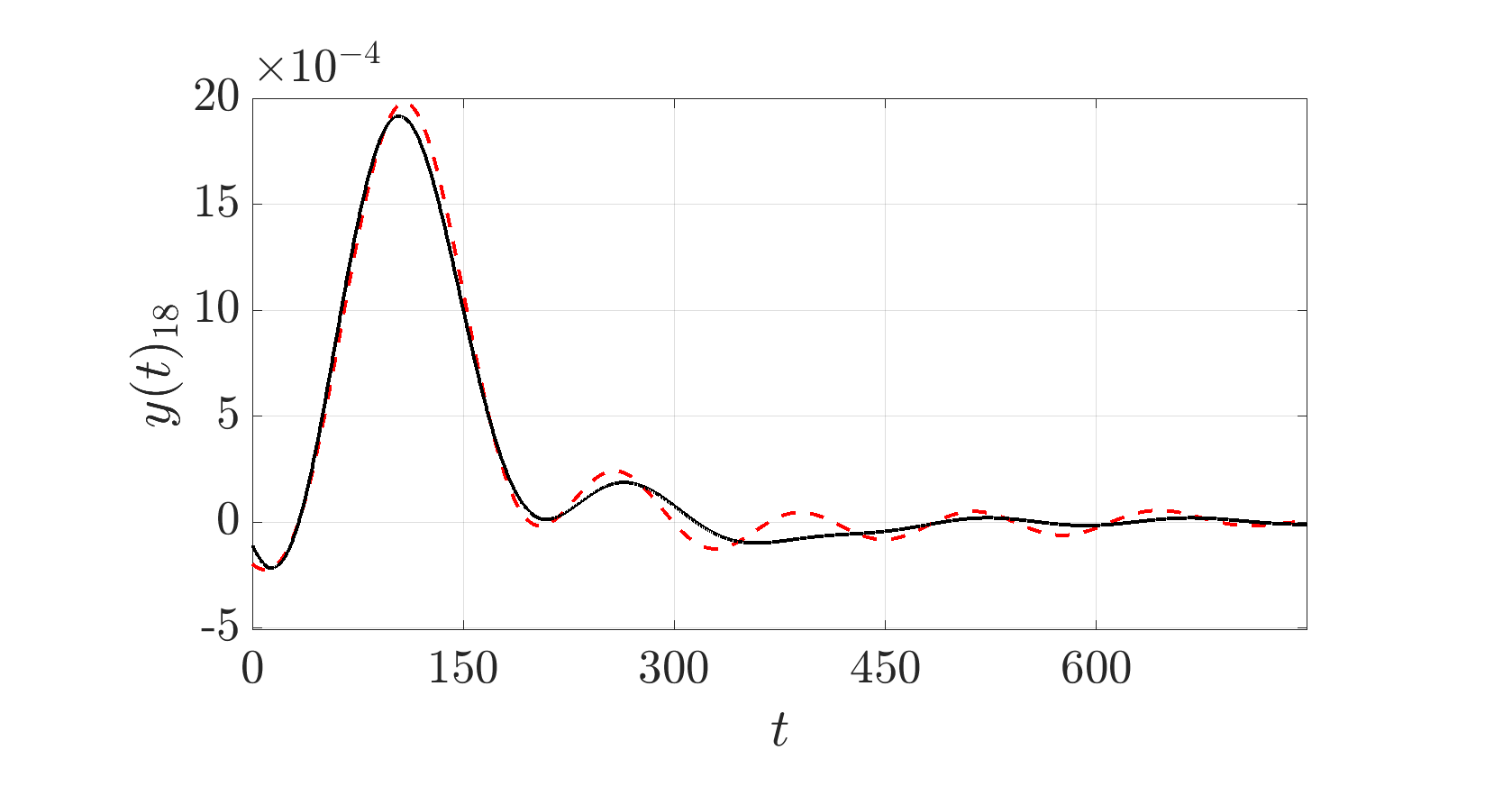}}%
{\includegraphics[width=0.50\textwidth,trim={50 45 120 85},clip]{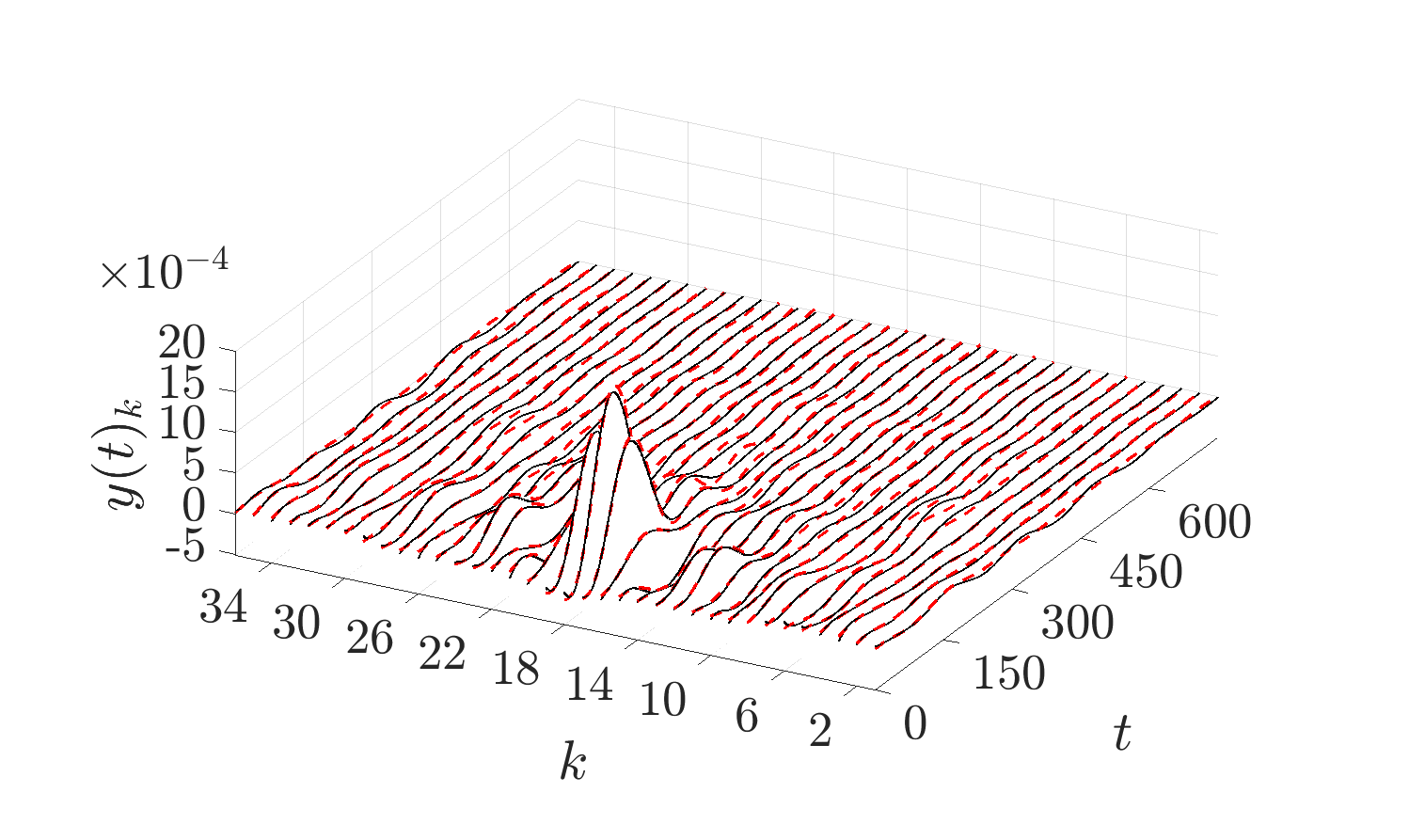}}\\
{\includegraphics[width=0.48\textwidth,trim={110 50 160 20},clip]{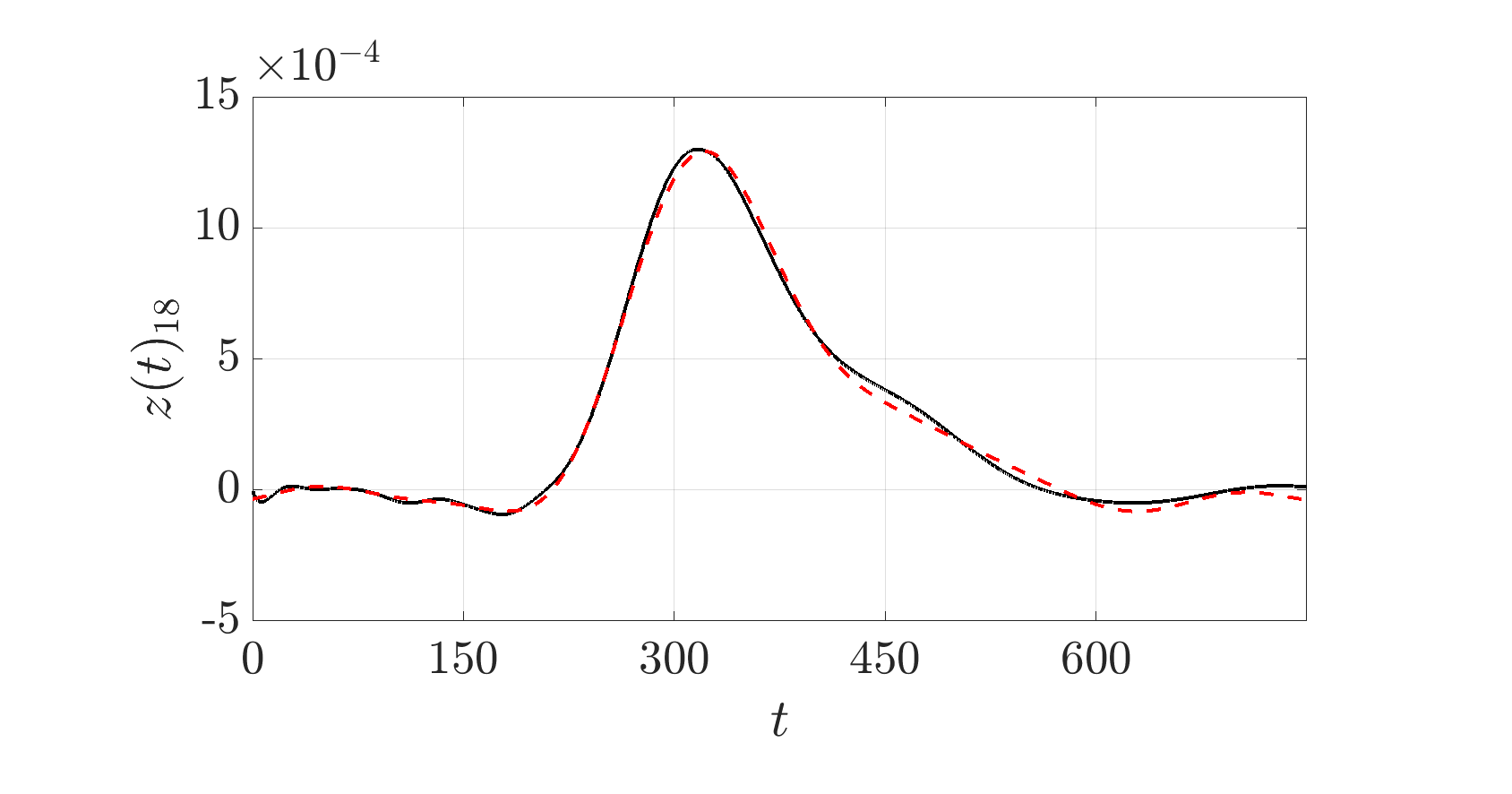}}%
{\includegraphics[width=0.50\textwidth,trim={50 45 120 85},clip]{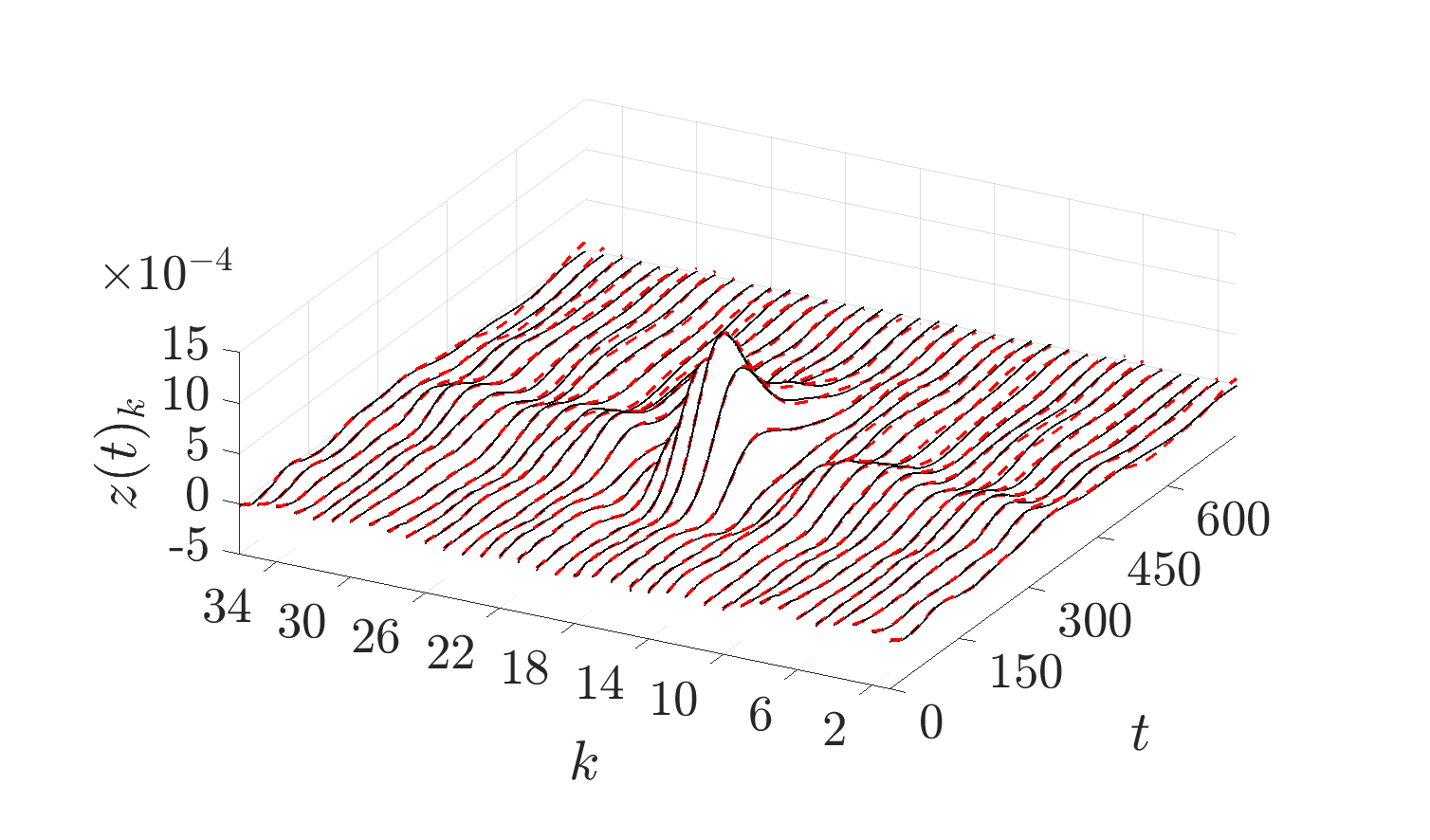}}\\
\end{center}
\caption{Original identified impulse responses used for the ERA (solid black lines) vs ROM impulse response (dashed red lines). Impulse response from $\mathbf{d}(t)$ to $\mathbf{y}(t)$ and $\mathbf{z}(t)$. The original identified impulse responses are built by the improved frequency response technique. Left: central output. Right: complete set of TF.}
\label{fig:eraTF}
\end{figure}

Given the position of the sensors, data-driven TFs can be computed by exploiting the methods outlined in \S~\ref{subsec:empiricaltf} and \S~\ref{subsec:era} and improved as described in \S~\ref{subsec:improvedtf}. Ensemble averaging is performed on time series with a sampling frequency of $1/0.3$, over a sampling time of $T=25000$, with a number 2500 of samples per each segment, an overlap of $80\%$, a total number of 166 segments, and by means of a triangular windowing function. The improvement of the empirical TF, with available outputs $y(t)_k$ at $x_1 = 250$ and estimated outputs $\tilde{z}(t)_k$ at $x_1 = 400$, is summarized in figure~\ref{fig:imprtf} and table~\ref{tab:imprtf}. Figure~\ref{fig:imprtf} shows that the empirical TF estimates an output $\tilde{z}(t)_k$ that is smoother than the original $z(t)_k$ taken from the DNS: the contours of the empirical TF do not show the sharp peaks of the DNS output (the darker areas). This implies that the empirical TF lacks accuracy in estimating the higher frequencies present in the original signal. The reason is likely spectral leakage together with the lower amplitude that the higher frequencies have with respect to the lower frequencies. This difference is decreased in the improved TF because the error, as in equation \eqref{eq:impTFerr00}, is computed in the physical domain, where it is possible to isolate the higher frequencies bypassing the issue of relative amplitude and spectral leakage. In fact, the improved TF estimates better the higher frequencies, as in figure~\ref{fig:imprtf}. Errors in amplitude estimation are also related to the windowing procedure of the time signal in the ensemble averaging. Even though the window is chosen to minimize such errors, its usage inevitably alters the computed amplitudes. Table~\ref{tab:imprtf} shows the normalized correlation value at zero delay, the mean-square ($\mathrm{MS}$) and the variance ($\mathrm{VAR}$) for the empirical TF and for the improved TF.\\ \indent
The TF that estimates the output $\tilde{z}(t)_k$ given the available output $y(t)_k$, i.e. $\hat{G}^{yz}$, is used in the IFFC control technique.
\begin{table}
\centering
\begin{tabular}{ccccc} 
                      & Corr. coeff.   & $\mathrm{MS}\left[ output \right]$ & $\mathrm{VAR}\left[output\right]$    \\ 
                      & at zero delay                  &                                                &                                                          \\ 
DNS data      & 1                                    & $2.82 \cdot 10^{-3}$               &$2.56 \cdot 10^{-3}$ \\
Empirical TF  & 0.90                               & $2.12 \cdot 10^{-3}$               &$1.91 \cdot 10^{-3}$                \\ 
Improved TF & 0.90         	                      & $2.44 \cdot 10^{-3}$               &$2.17 \cdot 10^{-3}$                \\ 
\end{tabular}
\caption{Correlation coefficient at zero delay (corresponding to its maximum value), mean-square $\mathrm{MS}[\mathbf{z}(t)]$ (DNS) or $\mathrm{MS}[\tilde{\mathbf{z}}(t)]$ (estimation), and variance $\mathrm{VAR}\left[\mathbf{z}(t)\right]$ (DNS) or $\mathrm{VAR}[\tilde{\mathbf{z}}(t)]$ (estimation) for the cases shown in figure~\ref{fig:imprtf}.}
\label{tab:imprtf}
\end{table}
The identified TFs used to mimic the effect of free-stream turbulence, characterized by $\mathbf{M}_d\mathbf{d}(t)$ in the ROM, assume as available output a set of sensors $y_d(t)_k$ at $x_1 = 175$ and estimate the outputs at $x_1 = 250,400$, i.e. $\mathbf{y}(t)$ and $\mathbf{z}(t)$. These TFs correspond to equation \eqref{eq:empiricaltf}.\\ \indent
The ROM resulting from the ERA consists of $N = 387$ degrees of freedom, which is considerably less than the degrees of freedom of the full system. The solution of the algebraic Riccati equations, which is the most computationally demanding step in the control design, with $N=387$ can be computed within the order of minutes nowadays (on a laptop). The value $N=387$ is found by imposing the error between the impulse response from the ROM and the original impulse response to be small enough. Since the ERA performs the Singular Value Decomposition of an Hankel matrix and the singular values are ordered such that $\sigma_{i} > \sigma_{i+1}$, the ratio $\sigma_N/\sigma_{max} \leq 5 \cdot 10^{-4}$ is used to determine $N$. Figure~\ref{fig:eraTF} compares the impulse response from $\mathbf{d}(t)$ in the ROM resulted from the ERA against the estimated TF used as original impulse response in the ERA. The TFs are centered in zero and present a peak which is related to the group velocity of the structures. There clearly is good agreement between the ROM and the original data.\par
%
%
\section{Control performance: transition delay}\label{sec:transdelay}
\begin{table}
\centering
\begin{tabular}{ccc} 
Control method   & Control problem          & Estimation problem           \\ 
IFFC                    & $Q = 1$,  $R = 2\cdot10^4$      & none                                                 \\ 
LQG                    &  $Q = 1$, $R = 50 $                   & $v_d = 1$, $v_n =5\cdot10^{-4}$  \\ 
\end{tabular}
\caption{Design weights of the $G^{yu}_m(t)$ used in the fully non-linear Navier-Stokes simulations.}
\label{tab:chosenpenalizations}
\end{table}
Here, the results from the non-linear simulations are presented.\par
\subsection{Transition delay}
The $G^{yu}_m(t)$ used in the non-linear N-S simulations are those resulting in the best performance in the control design. The weight matrices of the $J$ functionals of equations \eqref{eq:functionalforlqg} and \eqref{eq:WCobj00} are summarized in table~\ref{tab:chosenpenalizations} for IFFC and LQG, respectively. The choice of weights for control design is performed through input-output simulations based on time signals stored from uncontrolled non-linear N-S simulations. The input-output simulations consists in a linear superposition of signals, which makes them computationally inexpensive, such that their usage for control design becomes convenient to determine appropriate weights (details can be found in Appendix \ref{app:design}).\\ \indent
In order to assess the performance of the controller, the following quantity is introduced
\begin{equation}\label{eq:Jplotquantity}
\mathcal{E} = \frac{J_{controlled}^M}{J_{uncontrolled}^M}, \quad \quad J^M = \frac{1}{T} \int_0^T \mathbf{z}(t)^T\mathbf{z}(t) \ \mathrm{d} t,
\end{equation}
which corresponds to the average behavior of the output to minimize.  $T$ is the total time of the simulation.\par
\begin{figure}
\begin{center}
{\includegraphics[width=0.98\textwidth,trim={55 0 60 0},clip]{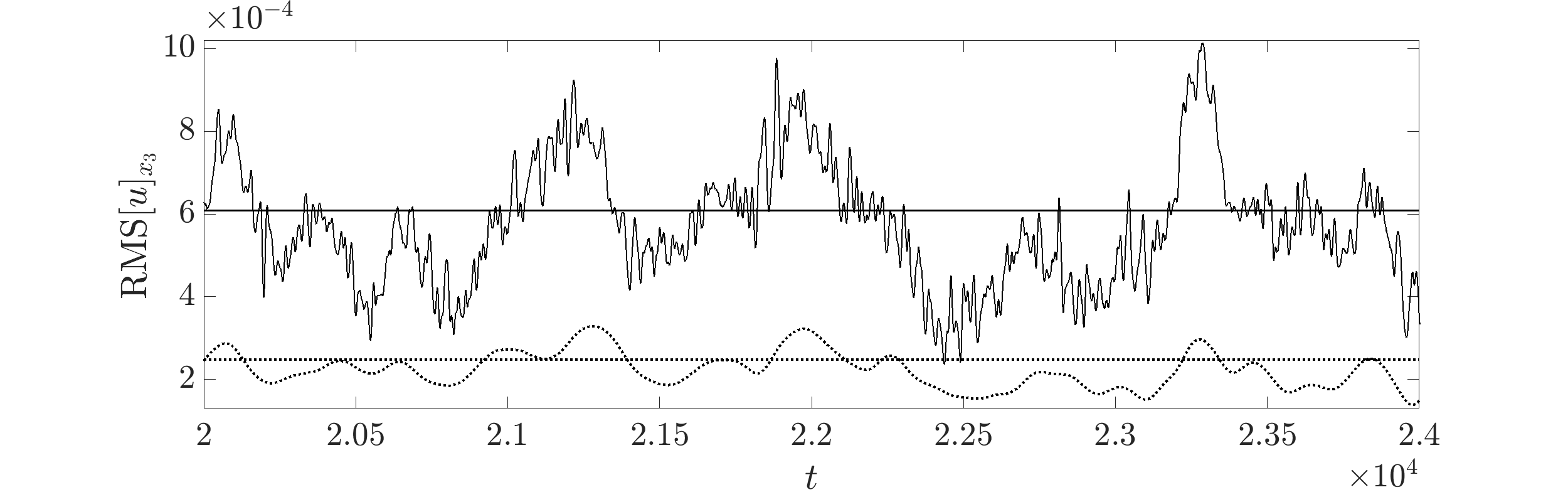}}\\
\end{center}
 \caption{Spanwise RMS values of the actuation signals $u(t)_{k}$ at $x_1 = 325$. Solid line: LQG. Dotted line: IFFC.}
\label{fig:actuation_signal}
\end{figure}
\begin{figure}
\begin{center}
{\includegraphics[width=0.98\textwidth,trim={35 0 60 0},clip]{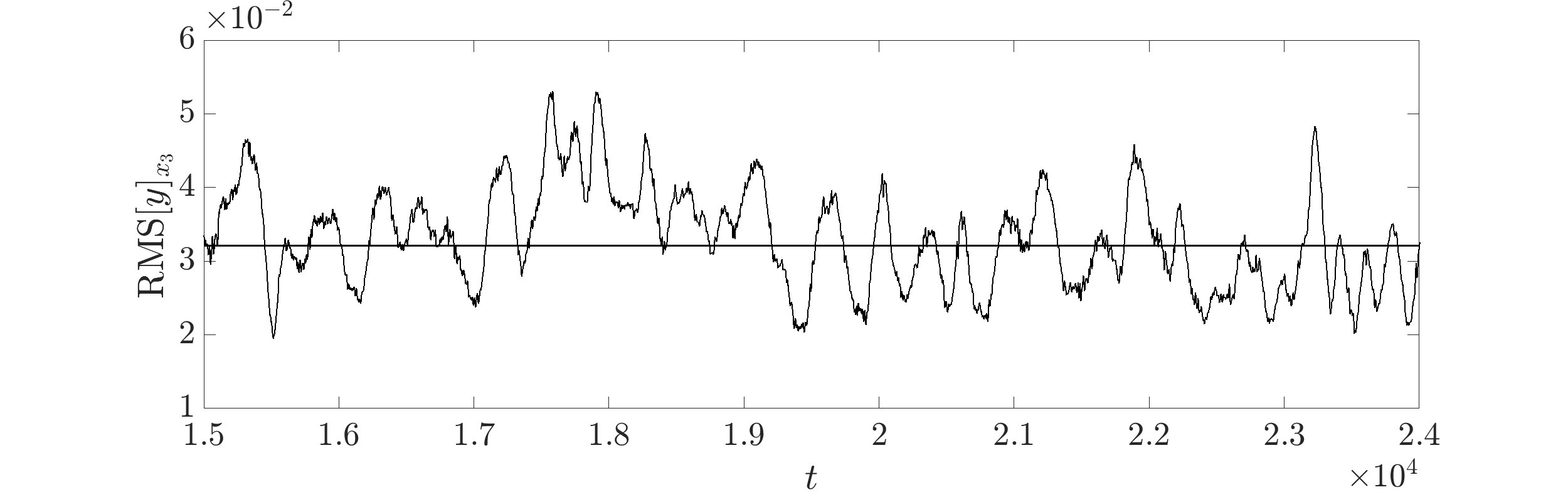}}\\
\end{center}
\caption{Spanwise RMS values of the output $y(t)_k$ at $x_1 = 250$.}
\label{fig:y_measure}
\end{figure}
\begin{figure}
\begin{center}
{\includegraphics[width=0.98\textwidth,trim={150 5 170 0},clip]{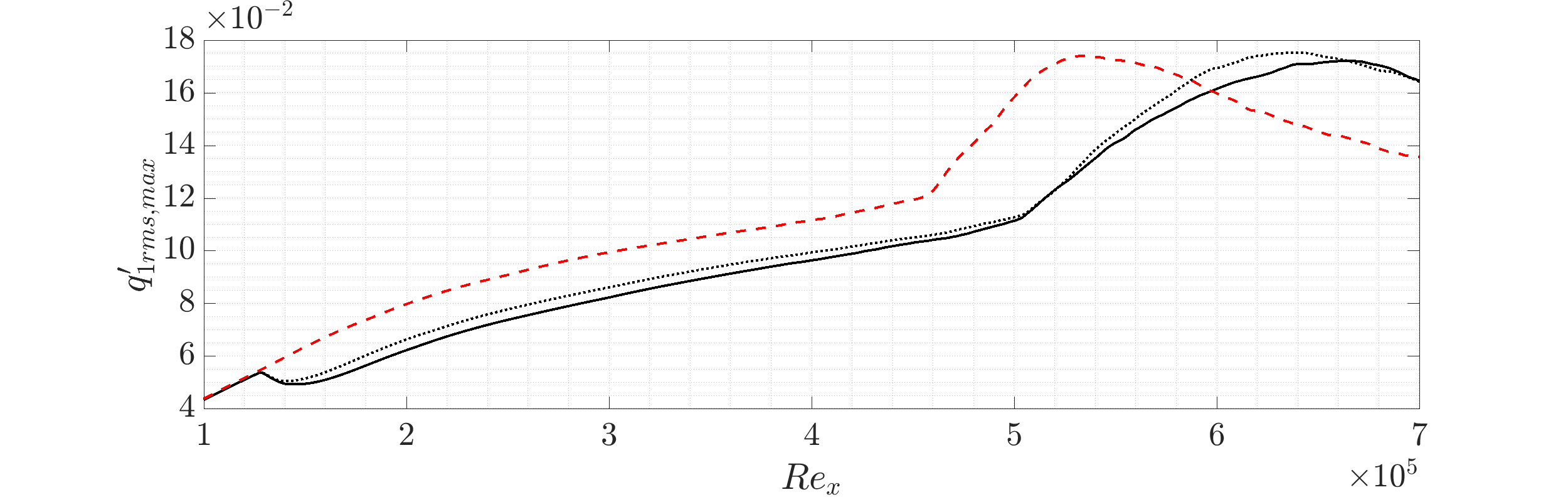}}\\
\end{center}
 \caption{Maximum along the wall-normal direction of the $q_{1,rms}$ averaged along the spanwise direction. Red dash-dotted line: uncontrolled case. Black dotted line: IFFC. Black solid line: LQG.}
\label{fig:umax}
\end{figure}
\begin{figure}%
\begin{center}
{\includegraphics[width=0.98\textwidth,trim={130 5 85 0},clip]{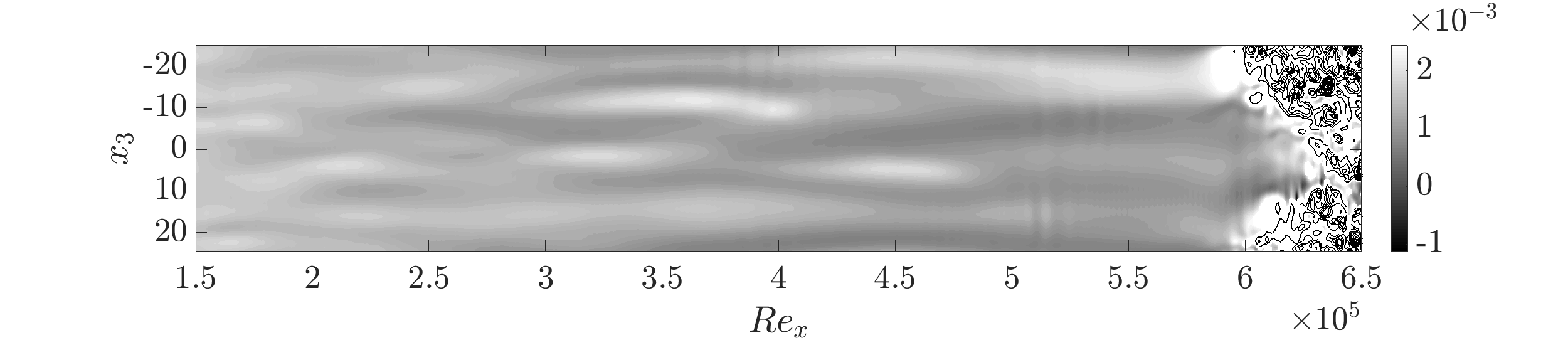}}\\
{\includegraphics[width=0.98\textwidth,trim={130 5 85 0},clip]{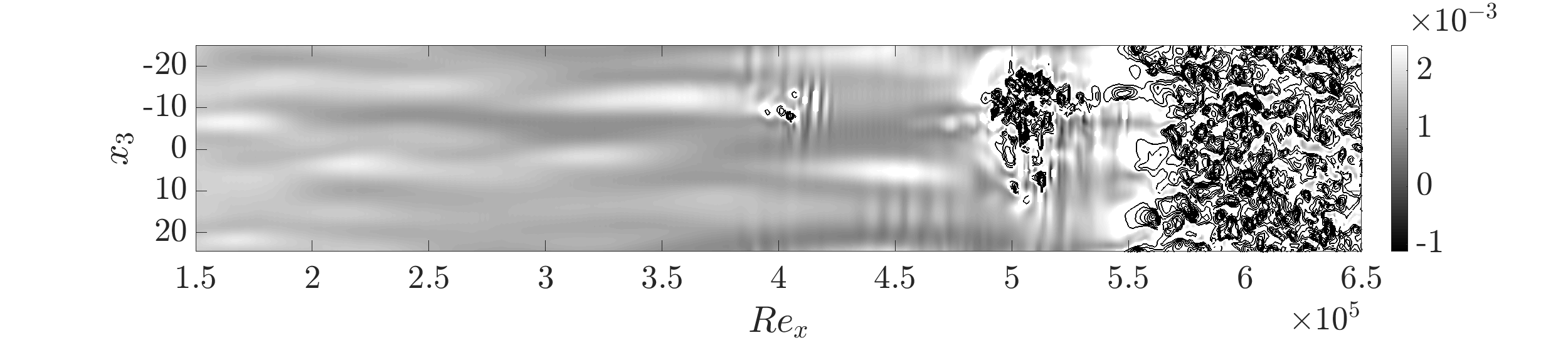}}\\
\end{center}
 \caption{Instantaneous skin friction coefficient $c_f(Re_x,x_3,t=t^*)$. Top: uncontrolled case. Bottom: controlled case with LQG. Black and white colors: $c_f < 2.5 \times 10^{-3}$. Empty contours: $c_f \geq 2.5 \times 10^{-3}$. Same time $t=t^*$ for both simulations; the starting seed for the random free-stream turbulence generation is the same.}
\label{fig:Cf2D}
\end{figure}

\begin{table}
\centering
\begin{tabular}{ccc} 
Control method             & Input-output sim.   & Non-linear sim.  \\ 
IFFC                 & 37.31 \%                                                                & 40.43\%             \\ 
LQG                  & 16.30 \%                                                                & 34.34\%            \\ 
\end{tabular}
\caption{Performance, $\mathcal{E}$, for the two control methods. Input-output vs non-linear N.-S. simulations.}
\label{tab:Joff_Jon}
\end{table}%
Table~\ref{tab:Joff_Jon} shows the performance of each control technique resulting from the input-output and non-linear N-S simulations. The comparison of the control techniques in the input-output simulation is consistent with the results of the fully non-linear Navier-Stokes: LQG outperforms IFFC. Moreover, both cases present a smaller reduction of the objective function in the results from the N-S simulations than the input-output ones, which may be attributed to non-linearities.
\begin{figure}%
\begin{center}
{\includegraphics[width=0.49\textwidth,trim={10 10 10 10},clip]{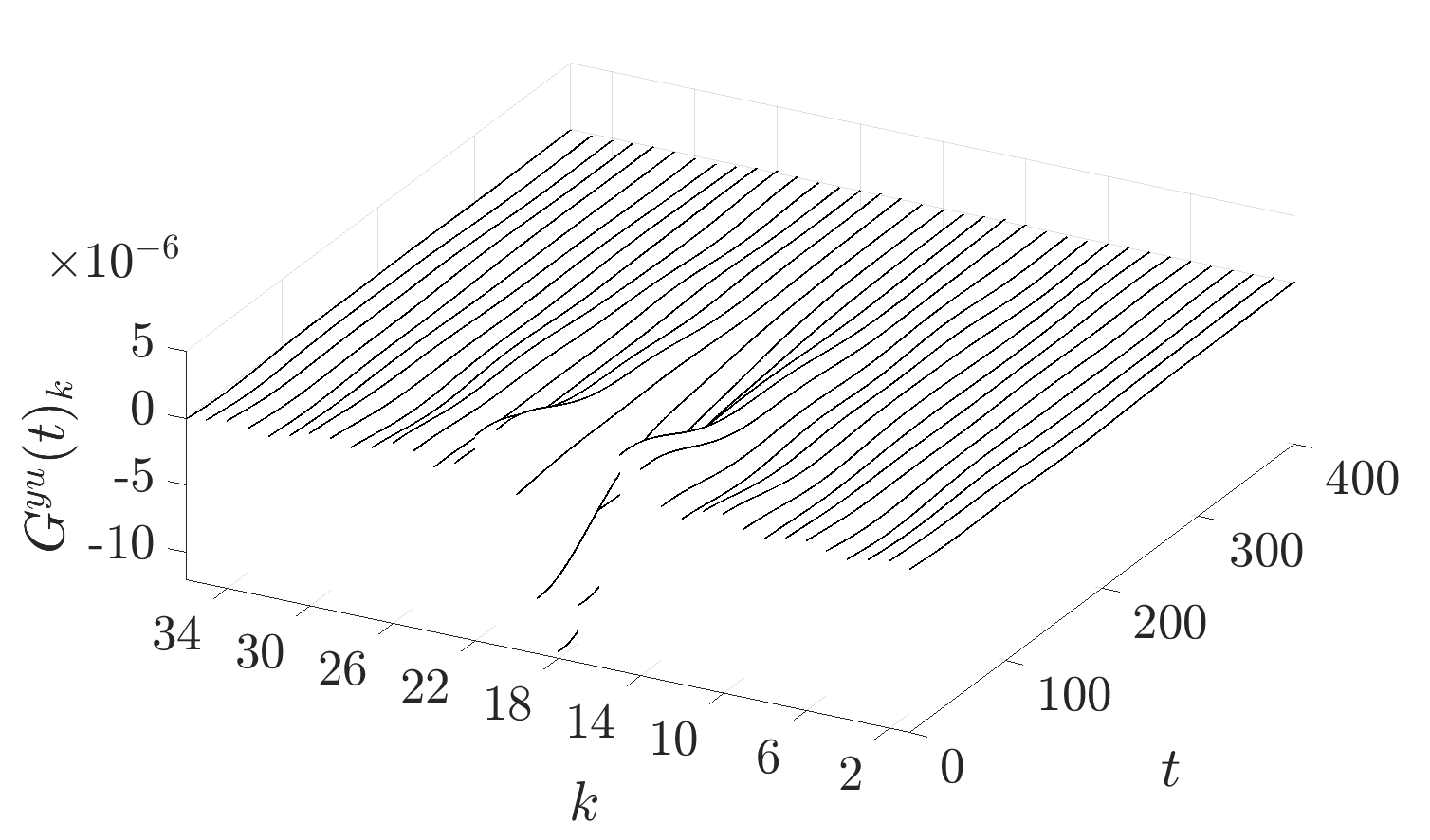}}
{\includegraphics[width=0.49\textwidth,trim={10 10 10 10},clip]{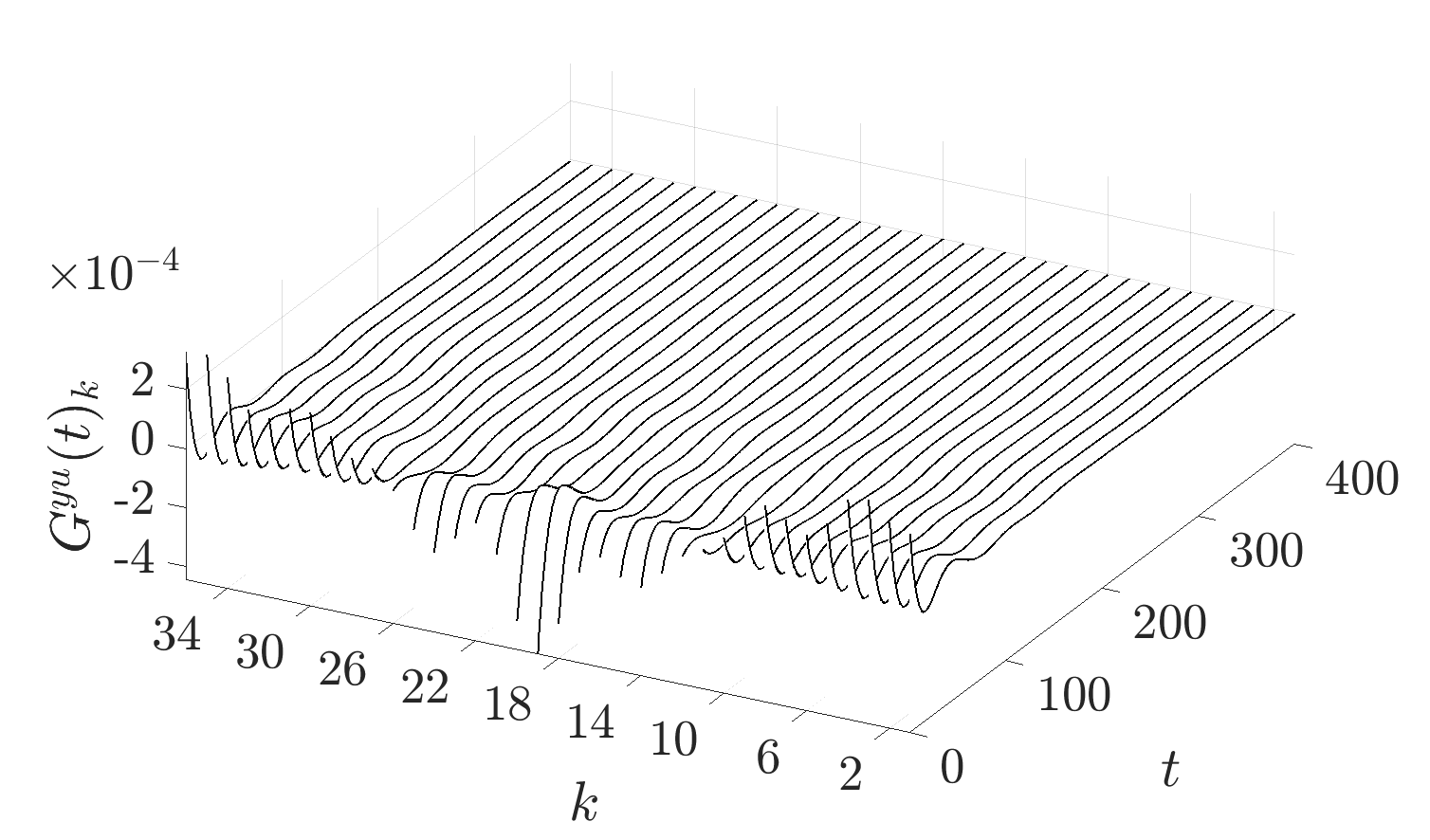}}\\
\end{center}
 \caption{Control TF $G^{yu}_k(t)$. Left: IFFC technique. Right: LQG technique. Notice the different scales for the gains between the two subfigures.}
\label{fig:kernels}
\end{figure}%
In figure~\ref{fig:kernels} the kernels from the IFFC and the LQG methods, respectively, are shown. These correspond to $G^{yu}_m(t)$ as in \S~\ref{sec:description_control}. It appears that the LQG weights a lot more the recent history of the signal than the IFFC does, which may be seen as the main reason for outperforming the IFFC result as further discussed next in \S~\ref{sec:causality}. In figure~\ref{fig:actuation_signal}, the spanwise RMS values of  $G^{yu}_m(t)$
\begin{equation}
\mathrm{RMS}[u]_{x_3} = \left( \frac{1}{N_u} \sum_{k=1}^{N_u} u(t)_{k}^2 \right)^{1/2} = \left( \frac{1}{N_u} \mathbf{u}(t)^T\mathbf{u}(t) \right)^{1/2},
\end{equation}
are plotted.
We observe that: (i) the magnitude of the averages and the fluctuation around the average values are higher for the signal generated by $G^{yu}_m(t)$ from the LQG, and (ii) the actuation signal from the IFFC is smoother. The first difference can explain why the performance of the LQG drops more than that of the IFFC when moving from the input-output to the Navier-Stokes simulations. In fact, the actuation signal multiplies a fixed spatial support, and an increase in the magnitude of the actuation signal corresponds to a more intense forcing that leads to stronger non-linearities. The second difference comes from the shape of the $G^{yu}_m(t)$ in figure~\ref{fig:kernels}. The $G^{yu}_m(t)$ from the IFFC is less localized around $t = 0$ than the one from the LQG, which means that the latter only captures low-frequency dynamics of $y(t)_k$ signals. This can also be seen by comparing figures \ref{fig:actuation_signal} and \ref{fig:y_measure}.\\ \indent
\begin{figure}%
\begin{center}
{\includegraphics[width=0.71\textwidth,trim={145 5 170 45},clip]{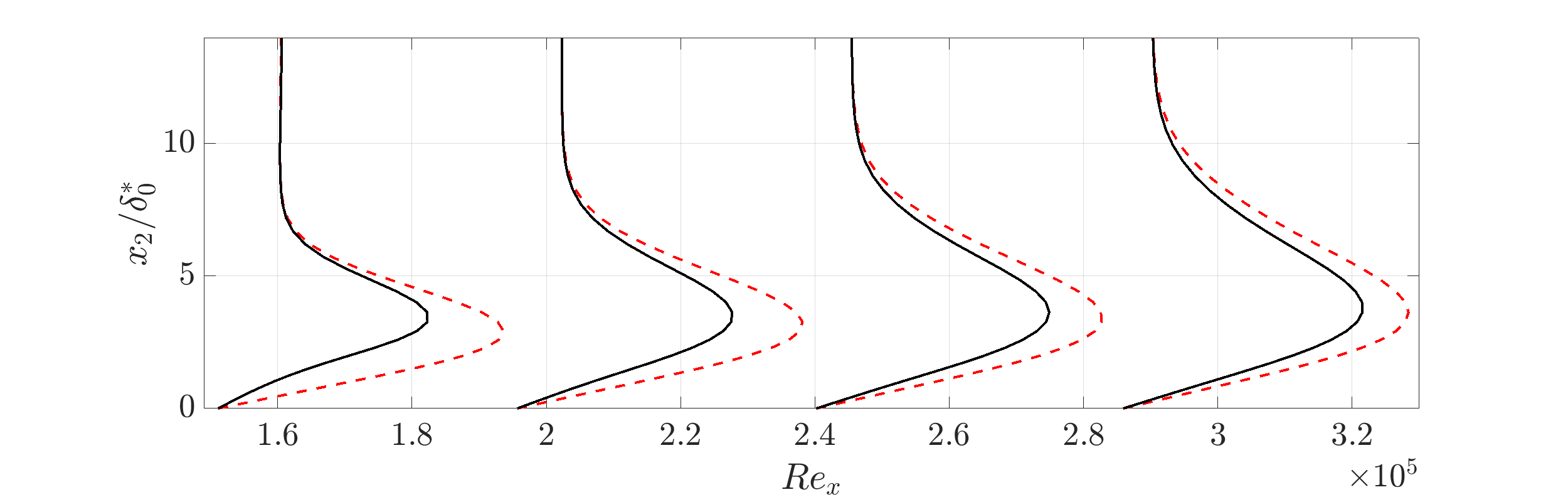}}
{\includegraphics[width=0.2735\textwidth,trim={7 3 62 25},clip]{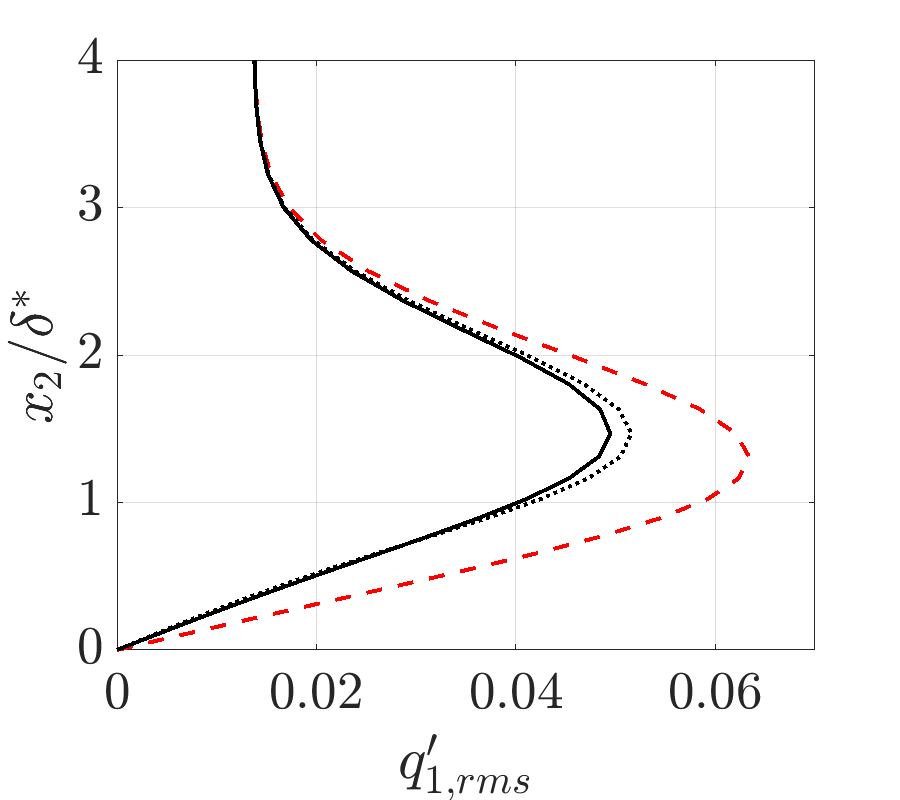}}
\end{center}
\caption{Left: $q'_{1,rms}/{q'}_{1,rms}^{uncontrolled}$ averaged along the spanwise direction at $Re_x = (1.51,1.96,2.40,2.86) \times 10^5$; red dash-dotted line: uncontrolled case; black solid line: LQG. Right: $q_{1,rms}$ averaged along the spanwise direction at $Re_x = 1.5\times 10^{5}$; red dash-dotted line: uncontrolled case; black dotted line: IFFC; black solid line: LQG.}
\label{fig:u_profile}
\end{figure}%
The effect of the actuation signal on the spanwise RMS values of the streamwise velocity component,
\begin{equation}\label{eq:urmsmax}%
q'_{1,rms} = \sqrt{< (q'_1 - <q'_{1}>_t)^2 >_{t,x_3}},
\end{equation}
with $< \bullet >$ representing the sample average, at $Re_x = (1.51,1.96,2.40,2.86) \times 10^5$, is shown in figure~\ref{fig:u_profile}. It is clear that LQG outperforms slighlty IFFC also in terms of reduction of the disturbance amplitude throughout the boundary layer. The disturbance energy drops by a factor of $\approx 40\%$ after the control action. As shown in figure~\ref{fig:umax}, despite the growth of disturbance amplitude beyond $x_1 = 400$ ($Re_x \approx 1.5 \times 10^5$) where the output $z(t)_k$ for the objective function is measured, a clear transition delay is achieved. \\
\begin{figure}%
\begin{center}
{\includegraphics[width=0.98\textwidth,trim={150 5 170 0},clip]{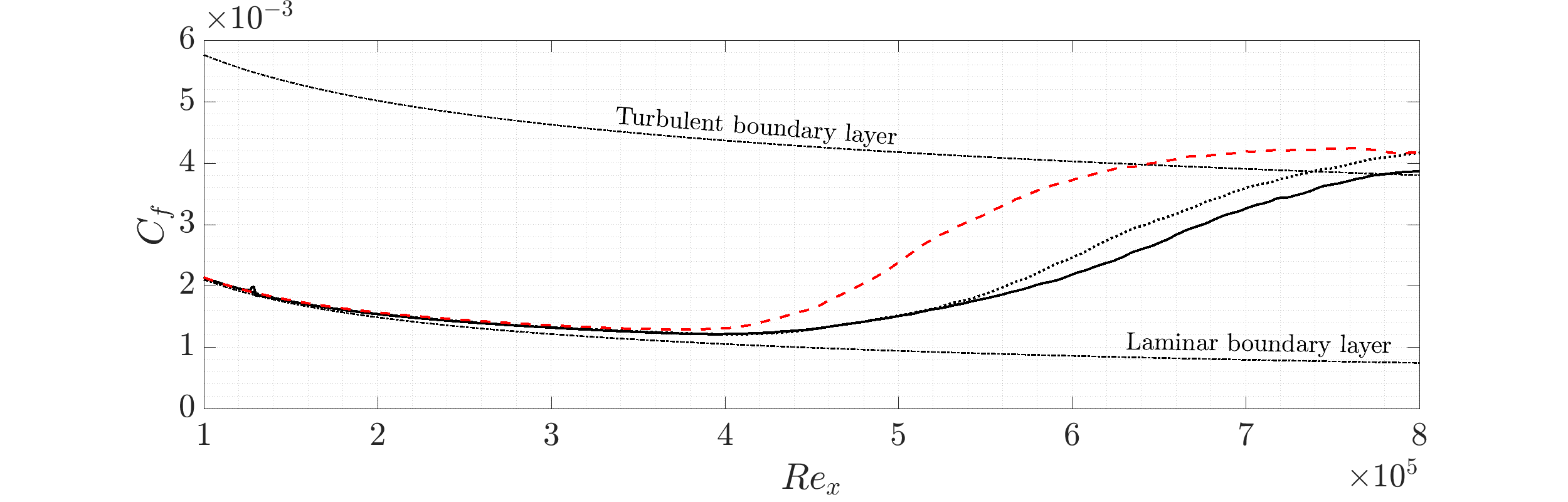}}
\end{center}
 \caption{Average skin friction coefficient $C_f$. Red dash-dotted line: uncontrolled case. Black dotted line: IFFC control method. Black solid line: LQG control method.}
\label{fig:cf}
\end{figure}%
A different measure of transition delay is the skin friction coefficient $C_f$, which explicitly appears in the computation of the drag and is the measure of interest for many applications. The behavior of the $C_f$ based on the RMS values of the streamwise velocity component,
\begin{equation}\label{eq:urmsmax}%
C_f(x_1) = \frac{\partial}{\partial x_2}  \left[ \sqrt{< {q'}_1^2 >_{t,x_3}}\right]_{x_2=0} ,
\end{equation}
is shown in figure~\ref{fig:cf}. There, the threshold curves that represents the skin friction of a laminar and a fully turbulent flat-plate boundary layer are also presented. It clearly appears again that LQG performs slightly better than IFFC and that in the best case the transition delay is around $\Delta Re_x = 1.5 \times 10^5 $, which is at least as good as most of the current results in literature where more idealized cases are studied, as in \cite{monokrousos2008a}. There a transition delay around $\Delta Re_x = 1.2 \times 10^5 $ is achieved in the best possible scenario for a case with turbulence intensity $Tu = 3.0\%$ and integral length scale $L = 5.0 \delta_0^*$. Nevertheless, in \cite{monokrousos2008a} actuation is performed by control of each point on a band of the flat plate, while measurements consists of wall shear-stress in both streamwise and spanwise directions and wall pressure fluctuations over a band of the flat-plate located downstream of the actuation.\\ \indent
A visualization of the instantaneous behavior of the skin friction coefficient over the flat plate,
\begin{equation}
c_f(x_1,x_3,t) = \frac{\partial}{\partial x_2}  \bigg[ q'_1(\mathbf{x},t) \bigg]_{x_2=0}
\end{equation}
is shown in figure~\ref{fig:Cf2D}. Although figure~\ref{fig:Cf2D} is a snapshot at an arbitrary $t = t^*$, it clearly appears that the controlled flow is smoother than the uncontrolled one for $Re_x < 5 \cdot 10^5$, it does not present the wiggles associated to secondary instability, and presents strong chaotic structures further downstream than the uncontrolled case, consistently with figure~\ref{fig:Cf2D}. These are all evidences of transition delay.\par
%
\subsection{Role of control methods, sensors and actuators for control performance}\label{sec:causality}
\begin{figure}%
\begin{center}
{\includegraphics[width=0.49\textwidth,trim={10 10 10 0},clip]{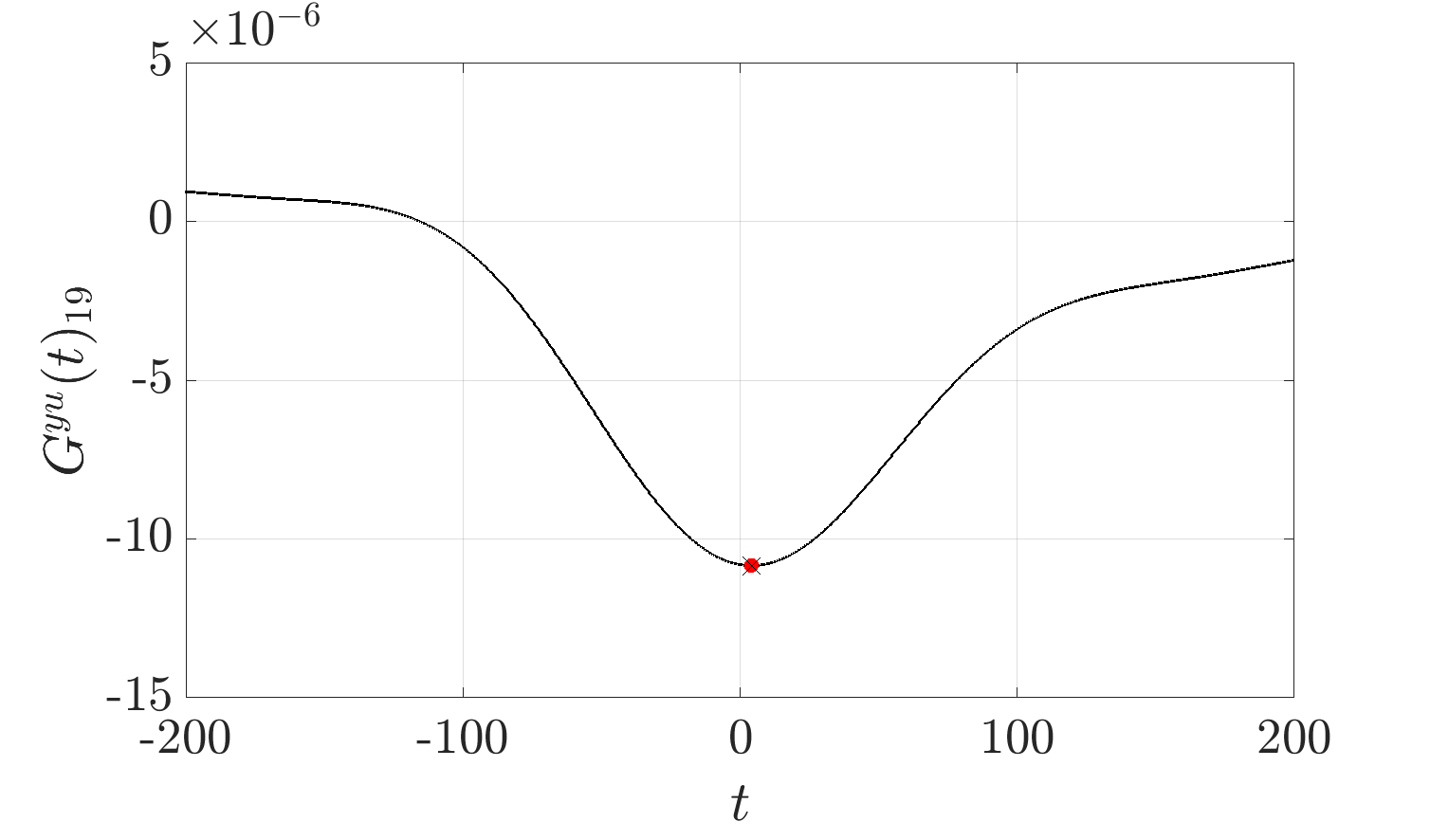}}
{\includegraphics[width=0.49\textwidth,trim={10 10 10 10},clip]{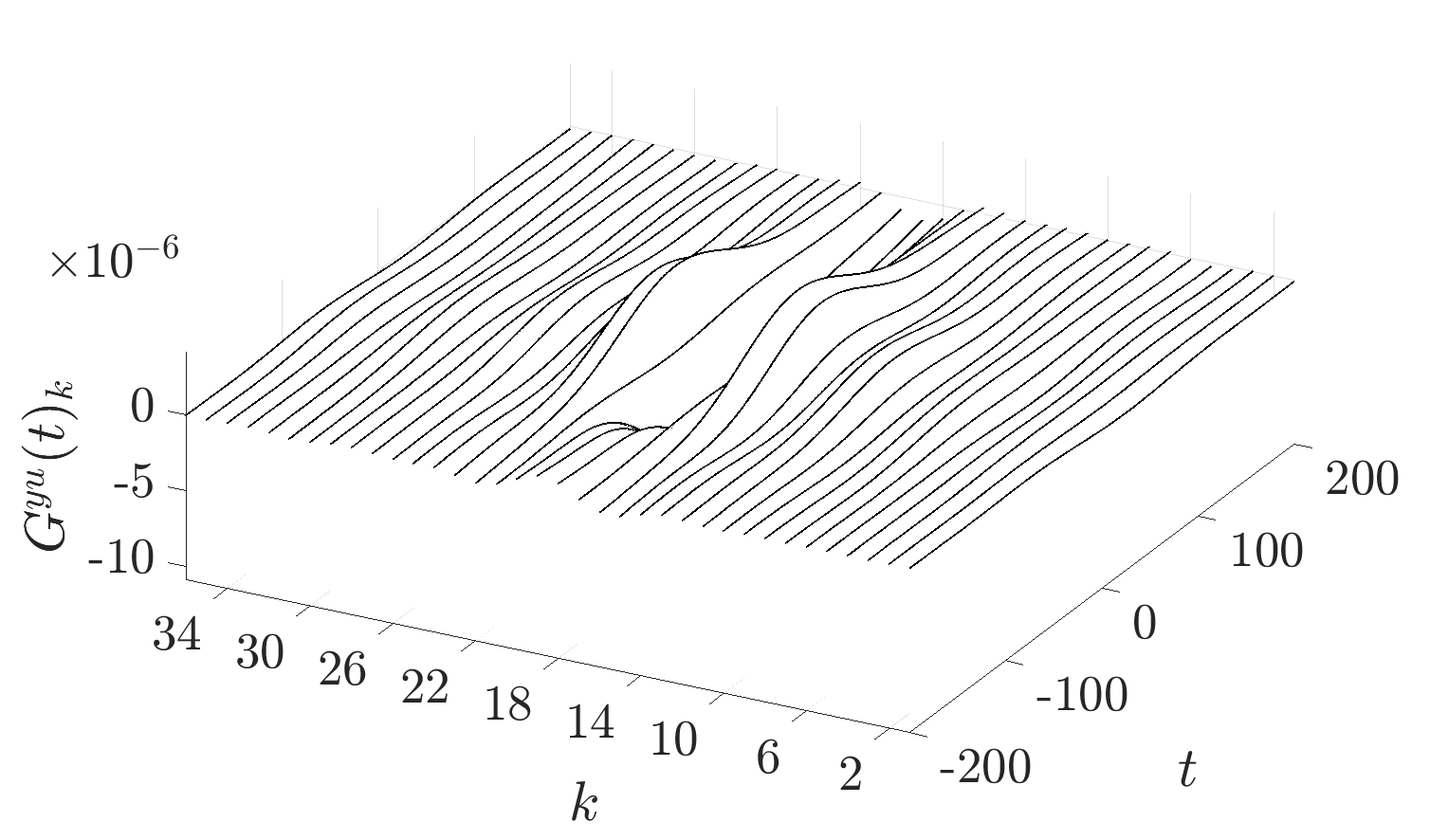}}\\
\end{center}
 \caption{Full non-causal control TF $G^{yu}_k(t)$ from the IFFC control method. Left: central line; the dot represents the peak value. Right: full TF.}
\label{fig:noncausalIFFC}
\end{figure}

\begin{figure}%
\begin{center}
{\includegraphics[width=0.309\textwidth,trim={5 10 20 30},clip]{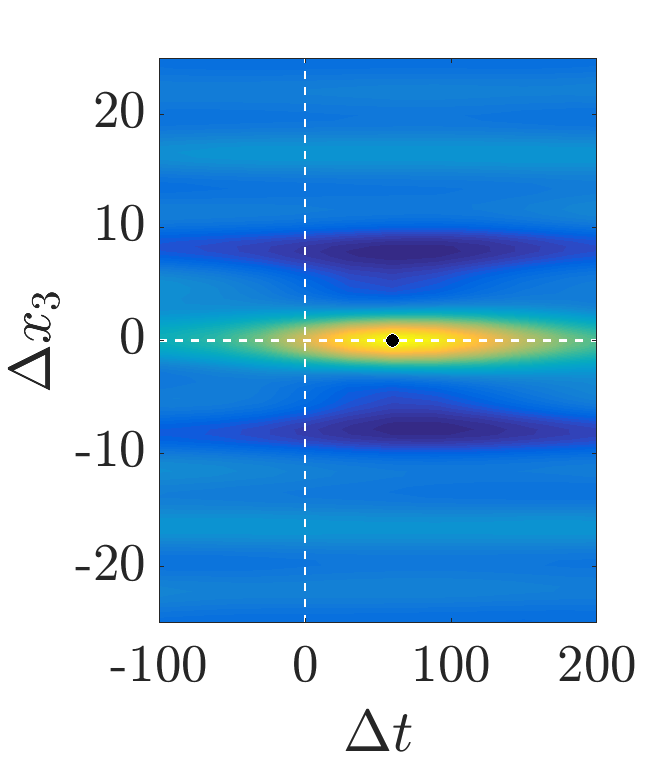}
\put(-117,118){$a)$}}
{\includegraphics[width=0.676\textwidth,trim={10 4 80 30},clip]{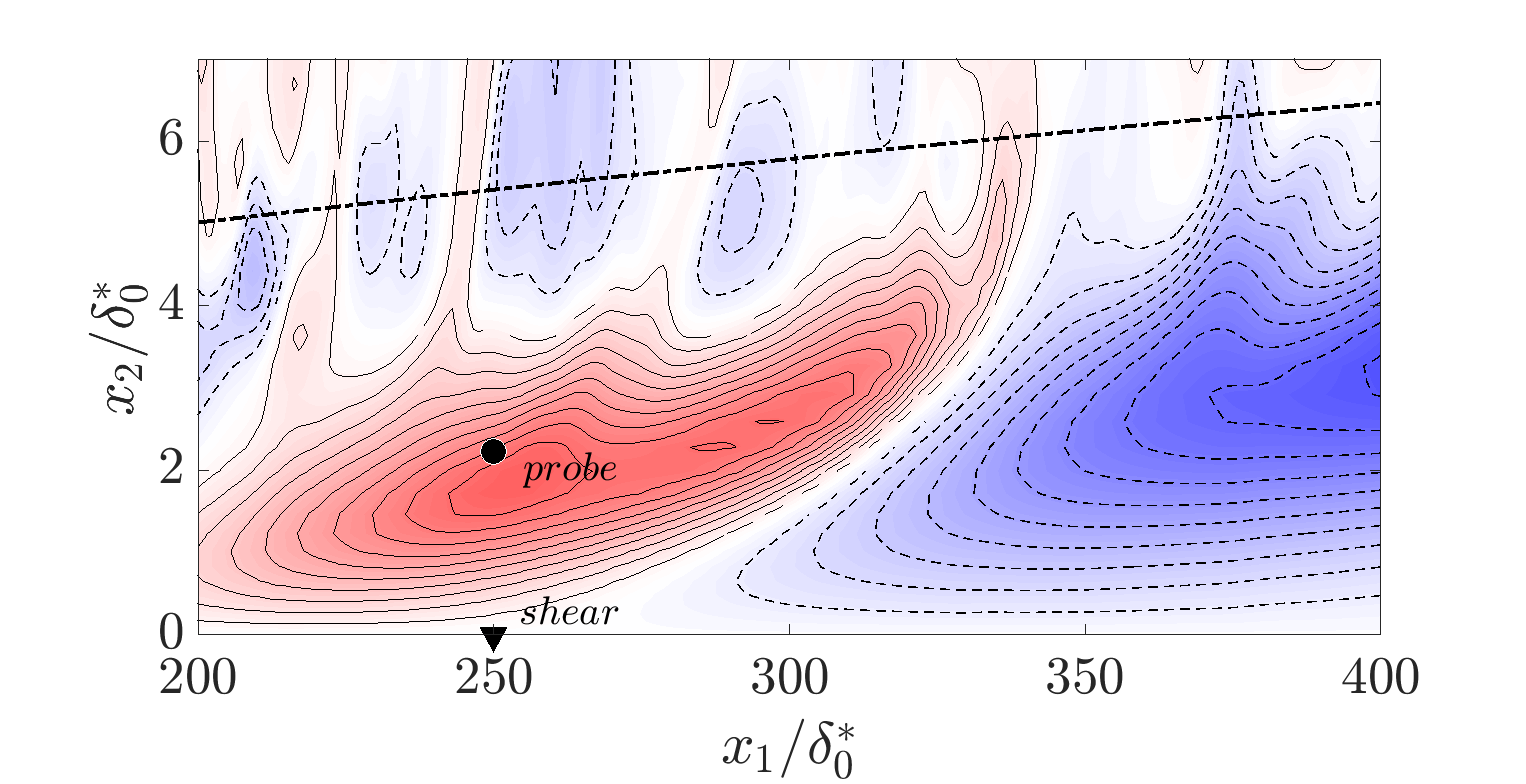}
\put(-244,118){$b)$}}
\end{center}
 \caption{Left, a) : cross-correlation between the streamwise velocity fluctuation at the peak of the $u_{rms}$ profile and the shear of the streamwise fluctuation at the wall. Right, b) : contours are the fluctuation of the streamwise velocity component; red and solid lines: positive values; blue and dashed lines: negative values; dashed-dotted line: $\delta_{99}/\delta_0^*$, boundary layer.}
\label{fig:correlation}
\end{figure}

The reason behind the LQG outperforming the IFFC is now discussed from a physical point of view.\\ \indent
A limiting characteristic for control performance is the relative position of sensors and actuators, which was chosen to minimize prediction errors and exploit the linear behavior of the flow field. The input-to-output and the output-to-output time delays are respectively $\tau_{uz} = 219.6$ and $\tau_{yz}^{IFFC} = 216$; these delays correspond to the peak of the transfer function, and approximate the average time for a structure induced by free-stream turbulence or the actuator, respectively, to arrive at the $\mathbf{z}(t)$ sensors. Therefore, the actuation is not sufficiently fast to cancel the streaks detected by the $\mathbf{y}(t)$ sensors, which is reflected by IFFC resulting in a non-causal $G^{yu}$, as shown in figure~\ref{fig:noncausalIFFC}. The result suggests that the performance of the controller may improve if it were possible to increase the difference between the time delays, which can be achieved either by changing the relative position of the devices or by changing the type of sensors or actuators. The limitations in the control performance are therefore not caused by the control methods, but by the structure of the plant. However, it appears that LQG can slightly compensate for this causality issue without any modification to the plant.\\ \indent
There exists a specific set of weights for which LQG outperforms IFFC, but there also exists a different set of weights for which LQG results in the same solution given by the IFFC. This is possible thanks to the presence of the estimation problem in the LQG, which introduces two more degrees of freedom in the control design. It follows that the reason behind the better performance of the LQG is the possibility of optimizing the estimation inside the control method, which is not included in the IFFC. Moreover, it appears that in the LQG keeping fixed the weight ratio $R^{LQG}/Q^{LQG}$ associated to the best solution and increasing the value of the ratio $v_n/v_d$ leads to worse performance (Appendix \ref{app:design}). Thus, since the weights of the estimation problem define the dynamics of the estimator (Appendix \ref{app:LQG}), the best solution comes from the estimator with the fastest response. In the present ROM the estimator that corresponds to $\tau_{yz}^{LQG} = \tau_{yz}^{IFFC} = 216$, as in the $G^{yz}$ used in IFFC, has a slower response than the one corresponding to the best solution found with LQG, where $\tau_{yz}^{LQG} = 213$. Thus, it appears that LQG achieves better performance for a case where the difference between the time delays, $\tau_{uz}-\tau_{yz}$, is higher than it is in IFFC, as suggested by the non-causal result found from the IFFC, and that better performance is achieved by an estimator with a fast response. The latter occurs because the weights of the estimation problem define the dynamics of its error: an estimator with a fast response has a fast decaying error. This implies that after the same $\Delta t$ the faster estimator is more accurate.\\ \indent
The connection between the shorter time delay $\tau_{yz}^{LQG} = 213$ mentioned earlier and the presence of a fast responding estimator may be explained by analyzing the capability of the wall shear-stress measurements to capture the dynamics of the streaks. An alternative measure of the streaky perturbations is their maximum streamwise velocity. Thus, a new set of outputs at $(x_1,x_2) = (250, 2.25)$ with the same spanwise positions of the other considered outputs is collected, as in figure~\ref{fig:correlation}b. These outputs are compared to those that measure the shear on the wall at $(x_1,x_2) = (250, 0)$, which are used to compute the input $u(t)_k$. Figure~\ref{fig:correlation}a shows the time-space correlation between the two sets of measurements. It appears that the output resulting from the measurements of the streamwise velocity perturbation and the output resulting from the measurement of the shear on the wall are highly correlated in the positive time half plane. With the considered convention for the correlation, this implies that fluctuations at the higher wall-normal position precede those on the wall, and thus the instantaneous measure of the shear on the wall cannot predict the instantaneous or future maximum velocity fluctuation of the streak. In other terms, there is reverse causality between the measurements at $x_2 = 0$ and the measurements at $x_2=2.25$. This effect can be associated to the tilting of the streaky structures shown in figure~\ref{fig:correlation}b: an advecting streak first passes at higher wall normal positions (exemplified by the considered probe), and only later leaves a wall shear-stress signature. However, the wall shear-stress measurements are not completely unable to estimate the streaks velocity; they can effectively predict the velocity of the tilted structure at wall-normal positions closer to the wall. There the velocity is lower, the convection velocity of the streaks is reduced, and thus the time delay $\tau_{yz}$, which describes their traveling time, is larger. A more accurate estimation, which corresponds to an estimator with a faster response, can effectively predict the velocity further from the wall. There the velocity is higher, so closer to the real traveling speed of the streaks, and the resulting time delay $\tau_{yz}$ smaller. This explains why to the best LQG solution corresponds a fast estimator and its connection to a shorter time delay $\tau_{yz}^{LQG} = 213$.\\ \indent
Moreover, the fact that the LQG results in a $G^{yu}$ which puts a lot of weight to the recent history of the output signal, around two orders of magnitude higher than the one from the IFFC (figure~\ref{fig:kernels}), is also explained by the presence of a fast estimator. In fact, to higher values of $v_n/v_d$, with $R_{LQG}/Q_{LQG}$ fixed to the value of the best solution, correspond a shape of $G^{yu}$ which approaches the one of the best IFFC solution, so the difference between the best LQG and IFFC results must come from the presence of the fast estimator. The last statement is consistent with the present discussion, as it implies that the performance of the controller improves when it mainly makes use of the portion of the output that lies in a small neighborhood of $t=T$, with $t \in \mathopen( {-\infty},T \mathclose]$ the history time and $T$ the running time. This neighborhood contains the meaningful information because of the mentioned reverse causality between the wall shear-stress measurements and the dynamics of the streaks, as shown by the non-positive time half plane in figure~\ref{fig:correlation}a.\\ \indent
Finally, it can be concluded that the performance of the best LQG result is better than those of the best IFFC thanks to the possibility of increasing the estimation accuracy through the estimation weights, which allows to slightly compensate for the reverse causality between the wall shear-stress measurements and the dynamics of the streaks. Moreover, it appears that the limitation caused by the structure of the plant, including the location of sensors and actuators and their shapes, is more critical than the choice of control technique, and thus is the key design challenge (as further discussed in the parallel work \cite{sasaki2019a}).\par
%
%
\section{Conclusions}\label{sec:concl}
The delay of bypass transition in a realistic scenario by means of active flow control, control theory and system identification is presented. Numerical simulations of the nonlinear transitional regime in a Blasius boundary layer are performed, where streaks are excited in the boundary layer by means of a high level of free-stream turbulence. A model-based method for the delay of bypass-transition realizable in experiments is introduced. It makes use of a ROM representation of the system and is based on the signals from a finite number of localized sensors and actuators placed on the wall, which mimic real shear-stress sensors and ring plasma actuators, respectively. A technique for the characterization of disturbances with a large number of degrees of freedom for model-based approaches is presented, which allows to obtain reasonably low-dimensional ROMs by isolating the dynamics of interest via system identification of the effects of the disturbance on the system. The method is reliable, easy to implement, and based on measurement data, in a data-driven approach that would be realizable in experiments. The presented technique is applied to generate the ROM via ERA for solving the flow control problem by means of LQG, which to the best of the authors' knowledge has never been done in a flow control application.\\ \indent
The performance of the LQG is compared to that of the IFFC optimal control technique, which does not need the explicit characterization of the disturbance on the system, thus, simplifying the flow control problem. LQG is seen to perform slightly better than the simpler IFFC method once appropriate weights in the cost function are selected. The performance of the control techniques are compared in linear input-output and non-linear Navier-Stokes simulations, showing that resorting to a linear ROM for control design is reasonable also in presence of the high-amplitude disturbances considered here. The effectiveness of the technique in delaying bypass transition is shown. Using LQG a transition delay of $\Delta Re_x \approx 1.4 \times 10^5$ for a case with turbulent intensity $Tu = 3.0\%$ and integral length scale $L = 7.5\delta_0^*$ is achieved. This highlights the capability of the presented methods to achieve at least as large delay of bypass transition as that obtained in more idealized cases found in literature \citep{monokrousos2008a}.\\ \indent
Finally, the differences in the results obtained with IFFC and LQG are analyzed and related to the structure of the plant, so the limitations caused by the relative positions of sensors and actuators and by the shape of the sensor are outlined. In particular, a reverse causality issue arising from using wall streamwise-shear-stress sensors to predict the dynamics of the streaks is shown. The way in which this causality issue limits the control performance is described, and an explanation on the way in which the LQG can compensate for such issue is provided.\par \bigskip
%
The authors would like to acknowledge the VINNOVA Projects PreLaFlowDes and SWE-DEMO and the Swedish-Brazilian Research and Innovation Centre CISB for funding. Moreover, part of this work was performed during an exchange programme at KTH, for which Kenzo Sasaki received a scholarship from Capes, project number 88881.132008/ 2016-01. Kenzo Sasaki work is also funded by a scholarship from FAPESP, grant number 2016/25187-4. Andr\'e V. G. Cavalieri was supported by a CNPq grant 310523/2017-6. The simulations were performed on resources provided by the Swedish National Infrastructure for Computing (SNIC) at NSC, HPC2N and PDC.
%
%
\appendix
\section{Linear Quadratic Gaussian regulator}\label{app:LQG}
The LQG technique is designed to solve the control problem on a dynamical system subject to stochastic white noise disturbances. Here, the dynamical system is a ROM and reads
\begin{equation}\label{eq:APPLQGsys00}
\begin{aligned}
\mathbf{\dot{q}}(t) &= \mathbf{A} \mathbf{q}(t) + \mathbf{B} \mathbf{u}(t) + \mathbf{M}_d \mathbf{d}(t),\\
\mathbf{y}(t) &= \mathbf{C}_y \mathbf{q}(t) + \mathbf{n}(t) ,\\
\mathbf{z}(t) &= \mathbf{C}_z \mathbf{q}(t).\\
\end{aligned}
\end{equation}
Since the LQG does not assume the full-state to be known, an estimation of the original dynamical system based on the known outputs is introduced,
\begin{equation}\label{eq:APPKFsys00}
\begin{aligned}
\dot{\tilde{\mathbf{q}}}(t)&=\mathbf{A}\tilde{\mathbf{q}}(t)+\mathbf{Bu}(t)-\mathbf{L}(\mathbf{y}(t)-\tilde{\mathbf{y}}(t)),\\
\tilde{\mathbf{y}}(t)&=\mathbf{C}_y \tilde{\mathbf{q}}(t),\\
\tilde{\mathbf{z}}(t)&=\mathbf{C}_z \tilde{\mathbf{q}}(t),\\
\end{aligned}
\end{equation}
where $\tilde{\mathbf{q}}(t)$ and $\tilde{\mathbf{y}}(t)$ are estimates of $\mathbf{q}(t)$ and $\mathbf{y}(t)$, and $\mathbf{L}$ is an $N \times N_y$ matrix to be designed. The estimated system accounts for the stochastic disturbances through the available outputs $\mathbf{y}(t)$.
Subtracting \eqref{eq:APPKFsys00} from \eqref{eq:APPLQGsys00} and substituting $\mathbf{y}(t)=\mathbf{C}_y\mathbf{q}(t)$ and $\tilde{\mathbf{y}}(t)=\mathbf{C}_y\tilde{\mathbf{q}}(t)$ gives
\begin{equation}\label{eq:APPerrdyn}
\dot{\mathbf{e}}(t) = ( \mathbf{A} + \mathbf{LC}_y ) \mathbf{e}(t) + \mathbf{M}_d \mathbf{d}(t) + \mathbf{L} \mathbf{n}(t),
\end{equation}
with $\mathbf{e}(t) = \mathbf{q}(t)-\mathbf{\tilde{q}}(t)$ the estimation error. Eq. \eqref{eq:APPerrdyn} shows that the error dynamics is based on the matrix $\mathbf{L}$ and is driven by the stochastic disturbances. Thus, the matrix $\mathbf{L}$ should stabilize the error dynamics and dampen the amplitude of the stochastic disturbance $\mathbf{n}(t)$.\\ \indent
Since $\mathbf{y}(t)$ is an available measure, the estimated system is deterministic. Its solution is used to compute the actuation input $\mathbf{u}(t) = \mathbf{K}\tilde{\mathbf{q}}(t)$, with $\mathbf{K}$ a matrix to be designed to solve the control problem. Substituting $\mathbf{u}(t) = \mathbf{K}\tilde{\mathbf{q}}(t)$ in \eqref{eq:APPKFsys00} gives
\begin{equation}
\begin{aligned}
\dot{\tilde{\mathbf{q}}}(t)&=(\mathbf{A}+\mathbf{BK})\tilde{\mathbf{q}}(t)-\mathbf{L}(\mathbf{y}(t)-\tilde{\mathbf{y}}(t)),\\
\tilde{\mathbf{y}}(t)&=\mathbf{C}_y \tilde{\mathbf{q}}(t),\\
\tilde{\mathbf{z}}(t)&=\mathbf{C}_z \tilde{\mathbf{q}}(t).
\end{aligned}
\end{equation}
The LQG technique consists in computing $\mathbf{K}$ and $\mathbf{L}$ to solve the control and the estimation problem, respectively. These two problems are usually coupled in optimal control, but in the LQG technique they are not and result in the minimization of two different $\mathcal{H}_2$-norms \citep{skogestad2005a}. The matrix $\mathbf{K}$ results from the linear quadratic regulator problem. It minimizes the objective function
\begin{equation}
\label{eq:APPobjlqr}
J = \lim_{T\to\infty} \frac{1}{T}\int_0^T (\mathbf{z}(t)^T \mathbf{Q} \mathbf{z}(t) + \mathbf{u}(t)^T \mathbf{Ru}(t)) \ \mathrm{d}t,
\end{equation}
and results in solving the following algebraic Riccati equation
\begin{equation}\label{eq:RiccatiLQ}
\mathbf{A}^{T}\mathbf{P}_u +\mathbf{P}_u\mathbf{A} - \mathbf{P}_u\mathbf{B}\mathbf{R}^{-1}\mathbf{B}^{T}\mathbf{P}_u + \mathbf{C}_z^{T}\mathbf{Q}\mathbf{C}_z = 0,
\end{equation}
where $\mathbf{P}_u$ is a positive semi-definite $N \times N$ matrix which is the unknown of the equation. The relationship between $\mathbf{K}$ and $\mathbf{P}_u$ reads
\begin{equation}
\mathbf{K} = -\mathbf{R}^{-1}\mathbf{B}^T\mathbf{P}_u.
\end{equation}
The matrix $\mathbf{L}$ results from the Kalman filter. It minimizes the expected value of the covariance matrix of the error at steady state,
\begin{equation}
\label{eq:functionalforestimator}
J = \lim_{t\to\infty} \mathrm{Tr}(\mathbf{P}_e(t)), \quad \quad  \mathbf{P}_e = \mathbb{E}\left[\mathbf{e}(t)\mathbf{e}(t)^{T}\right],
\end{equation}
with $\mathrm{Tr}(\bullet)$ the \textit{trace} operator, and results in solving the following algebraic Riccati equation
\begin{equation}\label{eq:RiccatiKF}
\mathbf{P}_e\mathbf{A}^{T} +\mathbf{A}\mathbf{P}_e - \mathbf{P}_e\mathbf{C}_y^{T}\mathbf{V}_n^{-1}\mathbf{C}_y\mathbf{P}_e + \mathbf{M}_d\mathbf{V}_d\mathbf{M}_d^{T} = 0,
\end{equation}
where $\mathbf{P}_e$ is a positive semi-definite $N \times N$ matrix and is the unknown of the equation, and $\mathbf{V}_d$ and $\mathbf{V}_n$ are the covariance matrices of $\mathbf{d}(t)$ and $\mathbf{n}(t)$, respectively. The relationship between $\mathbf{L}$ and $\mathbf{P}_e$ reads
\begin{equation}
\mathbf{L} = -\mathbf{P}_e \mathbf{C}_y^T\mathbf{V}_n^{-1}.
\end{equation}
Once both $\mathbf{K}$ and $\mathbf{L}$ are computed the state-space system based on $\tilde{\mathbf{q}}$ gives the input signal based on the history of the available output $\mathbf{y}(t)$,
\begin{equation}\label{eq:Gker_app}
\mathbf{u}(t) = -\int_0^{t}\mathbf{K}e^{(\mathbf{A+BK+LC}_y)(t-\tau)}\mathbf{Ly}(\tau) \ \mathrm{d} \tau.
\end{equation}\par
%
%
\section{Prediction for different streamwise positions}\label{app:pos}
\begin{figure}
\centering
\includegraphics[width=0.95\textwidth,trim={140 0 140 0},clip]{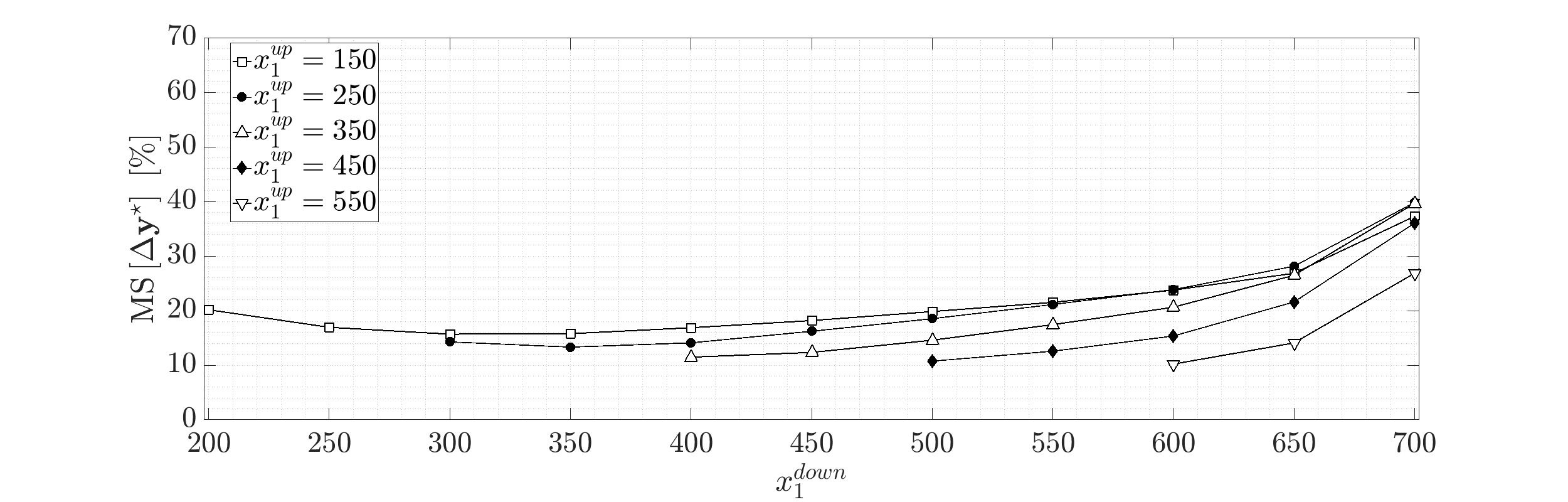}
\caption{Performance of the estimation of downstream outputs from available upstream outputs. $x_1^{up}$ is represents the available output position; $x_1^{down}$ represent the position of the estimated output. The outputs are the shear on the wall.}
\label{fig:accuracyofthepredictions}
\end{figure}
\begin{figure}
\centering
\subfigure{\includegraphics[width=0.49\textwidth,trim={5 15 10 15},clip]{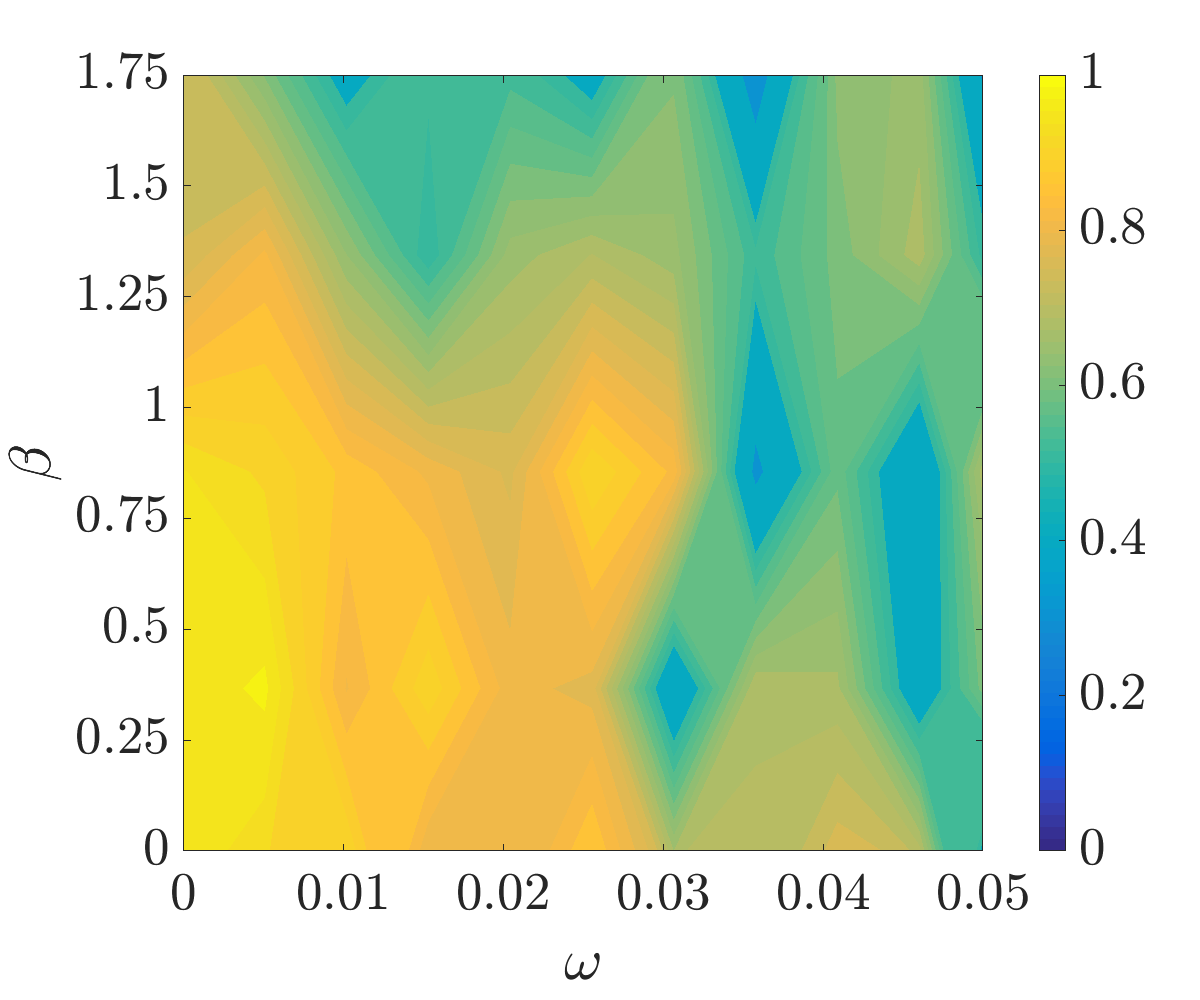}}
\subfigure{\includegraphics[width=0.49\textwidth,trim={5 15 10 15},clip]{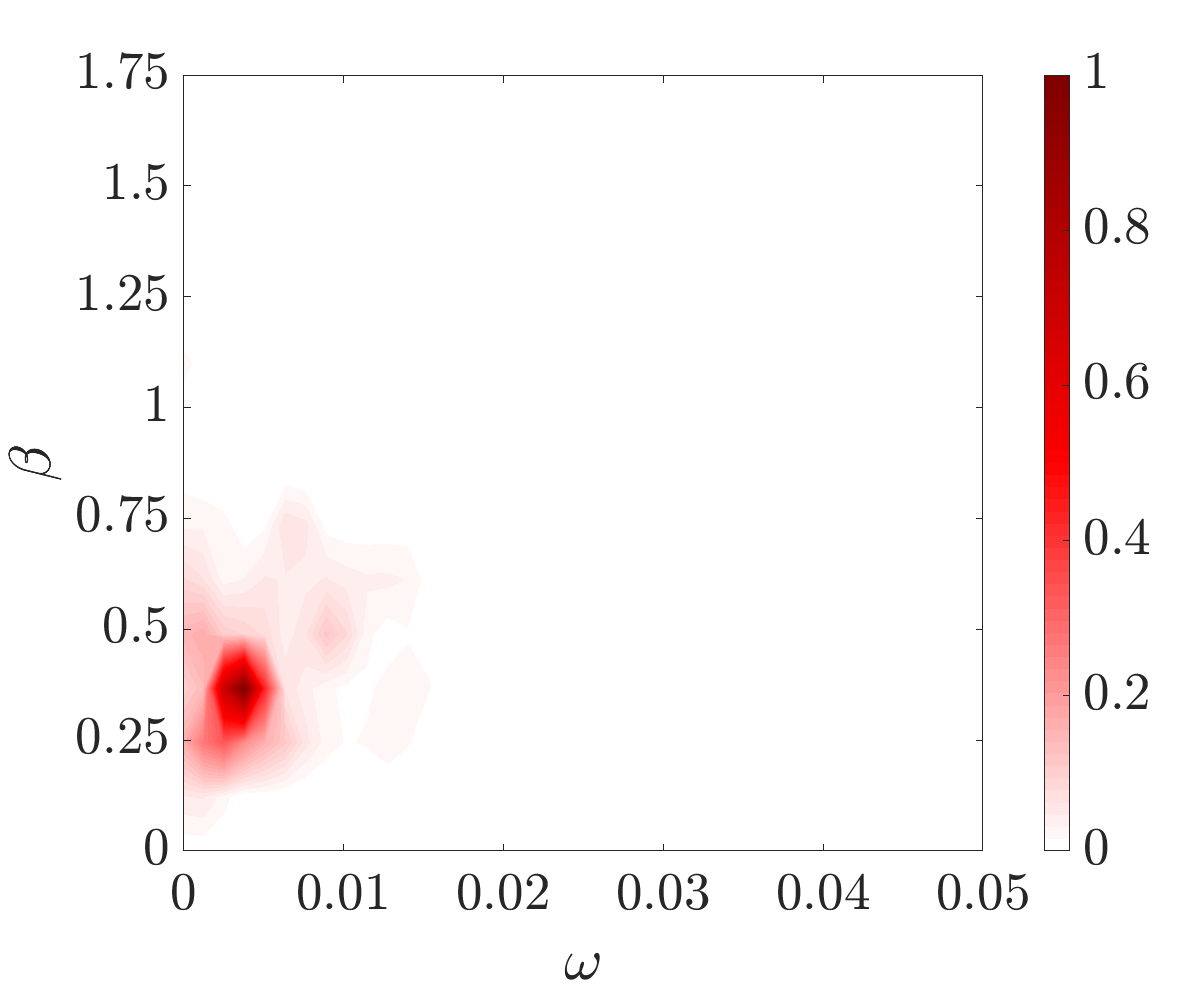}} \\
\caption{Left: Coherence $\gamma_{yz}$ between the outputs $y(t)_k$, at $x_1=250$, and $z(t)_k$, at $x_1=400$. Right: normalized power spectral density of the output $z(t)_k$: $\hat{S}_{zz}/\max(\hat{S}_{zz})$.}
\label{fig:coherencebetweendefinedinputandoutput}
\end{figure}

The streamwise position of sensors and actuators on the flat plate was chosen by also taking into account the estimation error. The behavior of the estimation error as function of the relative streamwise position between the available output and the output to estimate and as function of the absolute position of the available output is inspected. Given the set of $N_y=36$ estimated outputs $\tilde{\mathbf{y}}^{\star}_{est}(t)$ and the set of true outputs $\mathbf{y}^{\star}_{DNS}(t)$, the error is defined as
$\Delta\mathbf{y}^{\star}(t) = \mathbf{y}^{\star}_{est}(t)-\mathbf{y}^{\star}_{DNS}(t)$. The available outputs are placed at a streamwise position $x_1 = x_1^{up}$ and the estimated outputs $\mathbf{y}^{\star}_{est}(t)$ are placed at a streamwise position $x_1 = x_1^{down}$. The true outputs $\mathbf{y}^{\star}_{DNS}(t)$ are on the same place of the available outputs. It also holds that $x_1^{down} > x_1^{up}$, so the position of the outputs to be estimated never coincides to that of the available outputs. The available outputs are used to predict the downstream outputs in the future. Figure~\ref{fig:accuracyofthepredictions} shows the $\mathrm{MS}[\Delta\mathbf{y}^{\star}(t)]$. Estimation is performed by means of empirical TFs because of their low computational cost. In figure~\ref{fig:accuracyofthepredictions} it is evident that the current positioning of sensors and actuators, $250 \leq x_1 \leq 400$, is adequate. For $x_1^{up} = 150,250$, the $\mathrm{MS}[\Delta\mathbf{y}^{\star}(t)]$ initially decreases due to decay of free-stream turbulence intensity along the streamwise direction. For $x_1^{down} >350$, the $\mathrm{MS}[\Delta\mathbf{y}^{\star}(t)]$ increases in all cases because from that position the nonlinear interactions become important. The value of $\mathrm{MS}[\Delta\mathbf{y}^{\star}(t)]$ grows faster with increasing $x_1^{down}$ as the flow nonlinearity increases.\\ \indent
In order to further confirm the fact that the chosen positioning of sensors and actuators is adequate, the coherence coefficient between the measurements at $x_1 = 250$ and $x_1=400$ is calculated. The coherence $\gamma_{yz}$ is defined as
\begin{equation}\label{eq:cohercoeff}
\gamma_{yz}^2=\frac{|\hat{S}_{yz}|^2}{\hat{S}_{yy}\hat{S}_{zz}},
\end{equation} 
and measures the linearity between two different streamwise positions for each frequency. The definition holds for any pair of outputs. Its value varies between zero and one and indicates a complete random behavior (zero) and an exactly linear behavior (one) between the two signals.\\ \indent
It is desirable to have the highest values of coherence in $(\omega,\beta_k)$ regions where the signals are most energetic. This may be evaluated by computing the power-spectral density (PSD). The coherence coefficient between signals at $x_1=250$ and 400 and its normalized PSD is presented in figure~\ref{fig:coherencebetweendefinedinputandoutput}. The results indicate an almost linear relation between signals at these two streamwise positions for part of the $(\omega,\beta_k)$ space that is of interest.
It is noticeable that the most of the energy is strongly localized at very low frequencies, $\omega \approx 0$, with a spanwise wavenumber $\beta \approx 0.4$, which corresponds to streaky motions with slow streamwise variation. This is the reason for the observed good accuracy of the estimation and justifies the choices of placements for sensors and actuators.\par
%
\section{Choice of weights}\label{app:design}
The LQG has two objective functions, one for the control problem and one for the estimation problem, whereas the IFFC has only one objective function, the one for the control problem. The objective function for the control problem requires the definition of the weight matrix on the output $\mathbf{z}(t)$ (or $\hat{z}$), $\mathbf{Q}$ (or $\hat{Q}$),  and the penalization matrix on the input $\mathbf{u}(t)$ (or $\hat{u}$), $\mathbf{R}$ (or $\hat{R}$), while the estimation problem requires the matrices that describe the covariance of the stochastic disturbance $\mathbf{d}$, $\mathbf{V}_d$, and the noise $\mathbf{n}(t)$, $\mathbf{V}_n$. The control problem deals with finding the function that given an output provides an input to minimize an objective function, while the estimation problem deals with finding the function that allows to minimize the error in the estimation. The weights introduce more degrees of freedom in the design problem, and are usually left as free parameters. In fact, there is not a universally acclaimed method to compute those weights and close the control design problem, such that they are usually chosen iteratively \citep{skogestad2005a}. Here, a brute force method is applied: a grid of arbitrarily chosen weights is used and a set of $G^{yu}_m(t)$ is computed by means of the two control techniques as in \S~\ref{sec:description_control}. The computed $G^{yu}_m(t)$ are tested in input-output simulations based on the linear superposition of the input and output time series only. The input-output simulations make use of the second in equation \eqref{eq:FIRsys00} to compute the effect of $G^{yu}_m(t)$ on the reference output to annihilate, $\mathbf{z}(t)$, and on equation \eqref{eq:y2u00} for the relationship between $\mathbf{u}(t)$ and $\mathbf{y}(t)$. The input-output simulation consists in computing for each time step,
\begin{equation}\label{eq:FIRsys01}
\begin{aligned}
&u(t)_k = \int_0^t \sum_{m=1}^{N_y} G^{yu}_{m}(t-\tau) y(\tau)_{m+k-1} \ \mathrm{d}\tau,\\
&z(t)_k = \int_0^t \sum_{m=1}^{N_u} G^{uz}_{m}(t-\tau) u(\tau)_{m+k-1} \ \mathrm{d}\tau + z^d(t)_k,\\
\end{aligned}
\end{equation}
with $y(t)_k$ and $z^d(t)_k$, at $x_1 = 250,400$ respectively, with the time series of outputs saved from the nonlinear uncontrolled Navier-Stokes simulations. The method is thus a simpler simulation of the control effect considering only a linear superposition of the open-loop output $z^d(t)$ and what would result from control action (via the transfer function $G_m^{uz}$). The $k$ index was dropped in $G^{uz}_{km}(t)$ from equation \eqref{eq:y2u00} because the actuators have all the same spatial support and the linearized system dynamics is instantaneously homogeneous along the spanwise direction $x_3$. $G^{uz}_{m}(t)$ is found from an impulse-response simulation of the linearized Navier-Stokes equations.\\ \indent
This method avoids the use of computationally demanding Navier-Stokes simulations and is reliable to identify the best $G^{yu}_m(t)$ and the associated weights. It also proves to be consistent with the results of the nonlinear N-S simulations. The time required to perform the input-output simulations is of the order of seconds on the average laptop.\\ \indent
\begin{figure}
\begin{center}
{\includegraphics[width=0.49\textwidth,trim={5 5 5 5},clip]{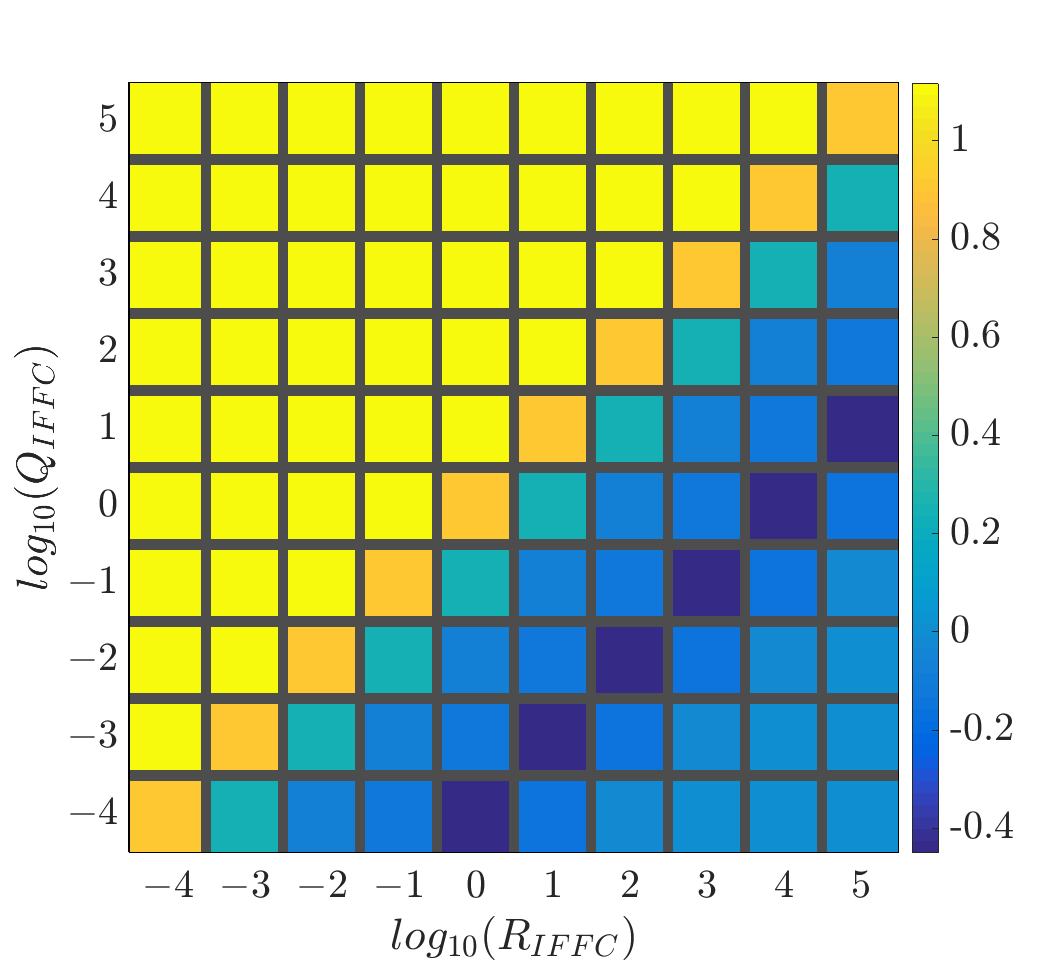}}
{\includegraphics[width=0.49\textwidth,trim={5 5 5 5},clip]{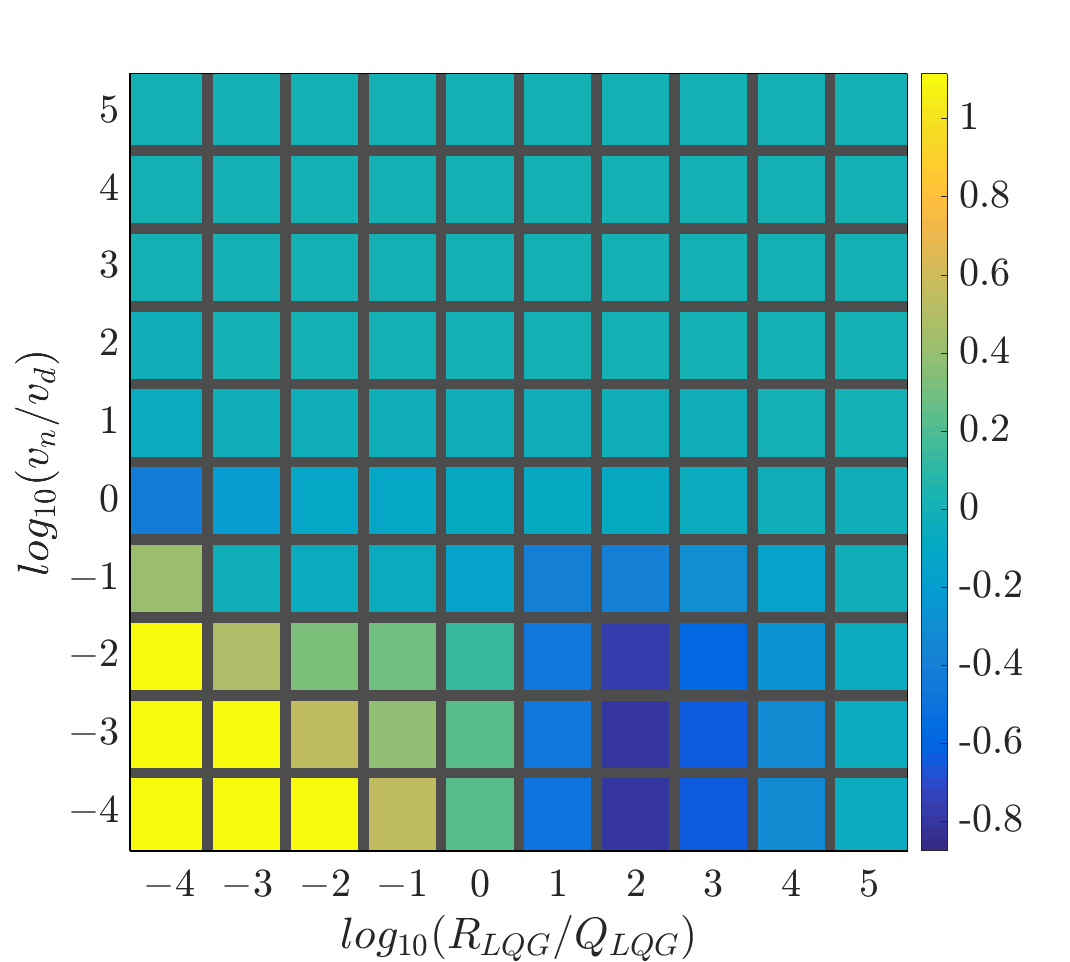}}
\end{center}
\caption{Performance parameter, $\log_{10}(\mathcal{E})$, as in equation \eqref{eq:Jplotquantity}, as function of the weight matrices. Left: IFFC control technique. Right: LQG control technique. }
\label{fig:Jmap}
\end{figure}%
Here, the covariance matrices are constants, as in equation \eqref{eq:covmat}, and the weight matrices for the control problem are chosen to be constants, 
\begin{equation}\label{eq:constweights}
\begin{aligned}
&\hat{Q} = Q_{IFFC},\\
&\mathbf{Q} = Q_{LQG}\mathbf{I},
\end{aligned}
 \quad \quad 
\begin{aligned}
&\hat{R} = R_{IFFC},\\
&\mathbf{R} = R_{LQG}\mathbf{I}.
\end{aligned}
\end{equation}
In figure~\ref{fig:Jmap} the results of the input-output simulations based on the IFFC or LQG methods are shown. From the results based on the IFFC technique it clearly appears that there exists a combination of weights $(Q,R)$ where $\mathcal{E}$ is constant. This occurs because the $J$ functional in equation \eqref{eq:WCobj00} can be written as $R$, which is a constant in this case, times another functional with only one weight in the form $Q/R$. The constant $R$ becomes irrelevant in the minimization problem, thus the minimization can be performed with respect to the functional with the weight $Q/R$. The weights can be related as
\begin{equation}\label{eq:QRfunc}
Q = c_1 R,
\end{equation}
with $c_1$ a constant. By using equation \eqref{eq:QRfunc}, the solution to the control problem based on the IFFC (equation \eqref{optimizedkernel}) can be written as
\begin{equation}
\hat{K}=\frac{\hat{G}^{H}_{uz}\hat{G}_{yz}}{c_1^{-1}+\hat{G}^{H}_{uz}\hat{G}_{uz}}.
\end{equation}
The results show (figure \ref{fig:Jmap}) $c_1 = 10^{-4}$. Moreover, by using equation \eqref{eq:QRfunc}, the objective function of the IFFC control technique (equation \eqref{eq:WCobj00}) can be written as
\begin{equation}\label{eq:newJIFFC}
J = R \int_{-\infty}^{\infty} \sum_k \left( c_1 \hat{u}^{H}\hat{u}+\hat{z}^{H} \hat{z} \right) \Delta \beta_k \ \mathrm{d}\omega ,
\end{equation}
which shows how under the assumption of constant weights the optimal solution depends only on the ratio of the weights. Writing the cost function as in equation \eqref{eq:newJIFFC} combines the physical meaning of the weights in one single parameter and constraints the solution of the minimization procedure to the isolines $J(Q,R) = constant$.\par
\begin{figure}
\begin{center}
{\includegraphics[width=0.48\textwidth,trim={110 50 160 20},clip]{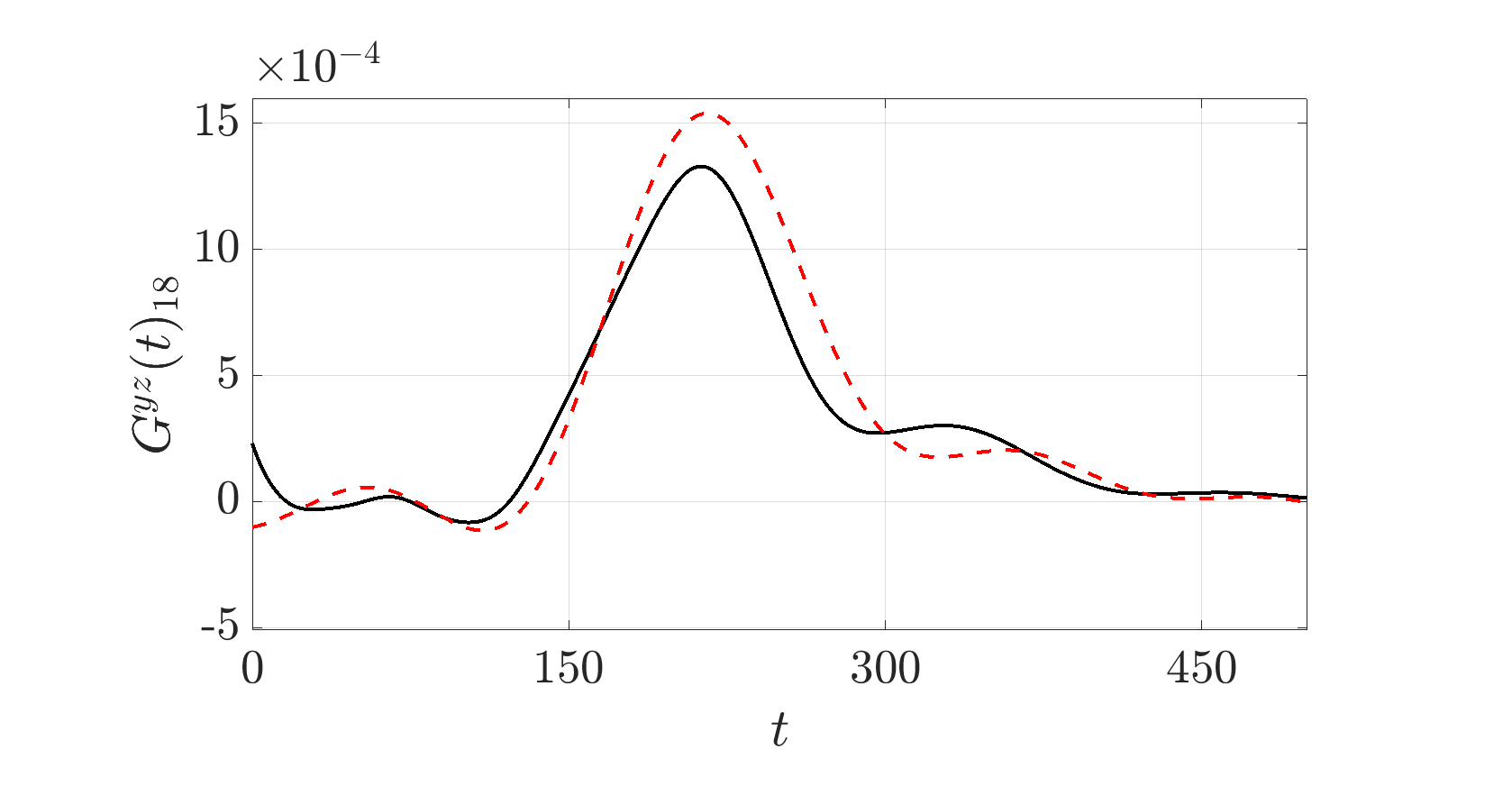}}%
{\includegraphics[width=0.50\textwidth,trim={50 45 102 85},clip]{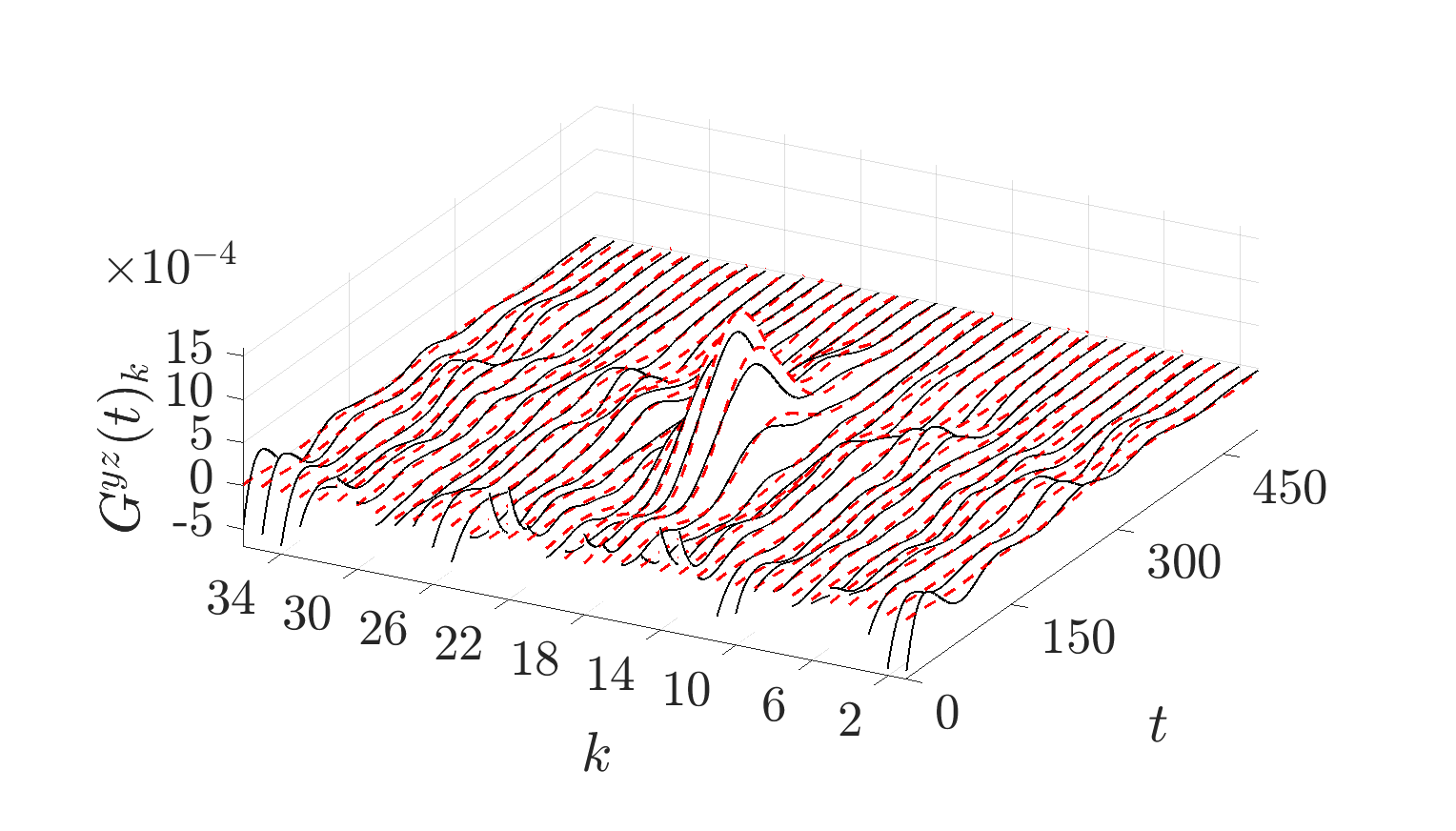}}\\
\end{center}
\caption{Estimation TF. Estimating $\tilde{\mathbf{z}}(t)$ from the available $\mathbf{y}(t)$. Solid black line: state observer. Dashed red line: empirical TF. Left: central TF. Right: complete TF.}
\label{fig:obsTF}
\end{figure}

Since the LQG results in two independent optimization problems, the control problem and the estimation problem (Appendix~\ref{app:LQG}), both the objective functions can be expressed in a similar fashion as in equation \eqref{eq:newJIFFC}. It follows that the performance $\mathcal{E}$ of the LQG can be expressed as function of two weight matrices only: one weight matrix from the control problem and one weight matrix from the estimation problem. This result is shown in figure~\ref{fig:Jmap} as function of $R_{LQG}/Q_{LQG}$ and $v_n/v_d$ (as in equation \eqref{eq:covmat}). Since the variables are associated to two independent optimization problems, there is no general reason for the existence of a set of weights for which the performance parameter $\mathcal{E}$ is constant and has a minimum. In fact, figure~\ref{fig:Jmap} shows that $\mathcal{E}$ has a minimum for a specific combination of $(R_{LQG}/Q_{LQG},v_n/v_d)$.\\ \indent
The minimum value achieved by the LQG is below the one achieved by the IFFC, $\mathcal{E}_{LQG}^{min} < \mathcal{E}_{IFFC}^{min}$. This occurs because the IFFC technique does not include the estimation problem. In figure~\ref{fig:obsTF} the estimation function of the LQG corresponding to $\mathcal{E}_{LQG}^{min}$ is compared to the $G^{yz}_m$, which is used in the IFFC and computed as in \S~\ref{subsec:improvedtf}. It can be seen that the two curves are slightly different. This occurs because the design parameters of the estimation problem can be tuned to seek for the overall optimal solution, which does not appear to be given by the IFFC, even though it was shown that $G^{yz}_m$ is a good estimation function. In other terms, since the LQG has more design parameters than the IFFC, it can span a space of solutions of higher dimensions than the IFFC. The latter statement also implies that there exist a combination of parameters for the estimation problem in the LQG for which $\mathcal{E}_{LQG} = \mathcal{E}_{IFFC}^{min}$ holds, and that the result achieved by the IFFC can be seen as a suboptimal solution of the LQG.\par
The input-output simulations allowed to identify the weights that give the best performance if the system were to be completely linear.\par
%
%
\bibliographystyle{jfm}
\bibliography{/Users/pmorra/Documents/KTH/Works/Bibliography/Morra_biblio}
\end{document}